\pdfoutput=1

\documentclass[11pt]{article}
\usepackage{jheppub}

\usepackage{amsmath}
\usepackage{amssymb}
\usepackage{graphicx,color,slashed,braket}
\usepackage{bbm}
\usepackage{ifpdf}
\usepackage{comment}

\usepackage{booktabs}
\usepackage{multirow}
\usepackage{makecell}
\usepackage{float}
\usepackage{verbatim}
\usepackage{caption}
\usepackage{cancel}
\usepackage{enumerate}
\usepackage{tcolorbox}
\usepackage[font=small,labelfont=bf]{caption}

  \usepackage{color}
  \definecolor{dark-gray}{gray}{0.20}
  \definecolor{gray}{gray}{0.30}
  \definecolor{light-gray}{gray}{0.80}
  \definecolor{dark-red}{rgb}{0.7,0,0}
  \definecolor{dark-green}{rgb}{0.1,0.4,0}
  \definecolor{dark-blue}{rgb}{0.3,0.3,0.7}
  \definecolor{light-blue}{rgb}{0.8,0.8,1}

\usepackage{hyperref}
\usepackage{setspace}
\hypersetup{
	colorlinks=true,
	linkcolor=dark-blue,
	citecolor=dark-red,
	urlcolor=dark-green,
	linktoc=page
}
\newcommand{\be}{\begin{equation}}
\newcommand{\ee}{\end{equation}}

\def\be{\begin{equation}}
\def\ee{\end{equation}}
\def\bea{\begin{eqnarray}}
\def\eea{\end{eqnarray}}

\newcommand{\nl}{\notag \\ &\quad\,}
\newcommand{\nll}{\notag \\ &}

\def\t{\tau}
\def\r{\rho}

\newcommand{\w}{\wedge}
\newcommand{\vol}{\text{vol}}

\renewcommand{\Im}{\text{Im}\,}
\renewcommand{\Re}{\text{Re}\,}
\renewcommand{\d}{\textrm{d}}
\newcommand{\rmi}{\textrm{i}}

\newcommand{\e}{\textrm{e}}

\newcommand{\dd}{\mathrm{d}}

\DeclareMathOperator{\sign}{sign}

\allowdisplaybreaks

\title{Scale-separated AdS$_4$ vacua of IIA orientifolds and M-theory}

\author[a]{N. Cribiori,}
\author[b]{D. Junghans,}
\author[c]{V. Van Hemelryck,}
\author[c,d]{T. Van Riet,}
\author[e,a]{T. Wrase}

\affiliation[a]{Institute for Theoretical Physics, TU Wien, Wiedner Hauptstrasse 8-10/136, A-1040 Vienna, Austria}
\affiliation[b]{Center of Mathematical Sciences and Applications, Harvard University, 20 Garden Street, Cambridge, MA 02138, USA}
\affiliation[c]{Instituut voor Theoretische Fysica, KU Leuven, Celestijnenlaan 200D, B-3001 Leuven, Belgium}
\affiliation[d]{Institutionen f\"{o}r Fysik och Astronomi, Box 803, SE-751 08 Uppsala, Sweden}
\affiliation[e]{Department of Physics, Lehigh University, 16 Memorial Drive East, Bethlehem, PA 18018, USA}

\emailAdd{niccolo.cribiori@tuwien.ac.at}
\emailAdd{djunghans@cmsa.fas.harvard.edu}
\emailAdd{vincent.vanhemelryck@kuleuven.be}
\emailAdd{thomas.vanriet@kuleuven.be}
\emailAdd{timm.wrase@lehigh.edu}

\notoc

\abstract{We revisit various aspects of AdS$_4$ flux vacua with scale separation in type II supergravity and M-theory. We show that massless IIA allows both weakly and strongly coupled solutions for which the classical orientifold backreaction can be tuned small. This is explicitly verified by computing the backreaction at leading order in perturbation theory. We give evidence that the strongly coupled solutions can be lifted to scale-separated and sourceless (but classically singular) geometries in 11D supergravity. 
}

\begin{document}

\begin{flushright}
UUITP-29/21
\end{flushright}

\maketitle

\newpage
\tableofcontents
\newpage
\section{Introduction}
The extra dimensions of critical string theory and M-theory are a blessing and a curse at the same time. They are a blessing since they provide a way to arrive at low-energy theories in 4D with many fields and interactions, starting from a unique theory in higher dimensions. They are a curse in the sense that we do not observe extra dimensions up to energies currently explored in accelerators. This means that one has to face an enormous hierarchy problem from the start: why are the visible dimensions at least of the Hubble scale and the compact dimensions of vastly smaller length scales?\footnote{Brane-world scenarios form an interesting alternative, which we do not discuss in this paper.}

Compactifications with a high degree of computational control exist when a sufficiently large amount of supersymmetry is preserved. In such compactifications to Minkowski space, the volume modulus of the compact dimensions can be a flat direction, also at the quantum level. In that case, it is not difficult to find solutions with arbitrarily small volume since the volume can be dialed at will. However, once supersymmetry is broken to $\mathcal{N}=1$ or even $\mathcal{N}=0$, this feature is lost. 
Indeed, particle physics and cosmology tell us that supersymmetry is broken at low energies. We therefore need to understand whether scale-separated vacua with little or no supersymmetry are possible in string theory. Once this is understood, we can turn to the question of how we ended up in such a vacuum. 

In this paper, we take a modest attitude and investigate to what extent \emph{Anti}-de Sitter (AdS) vacua with stabilised modes, $\mathcal{N}\leq 1$ supersymmetry and scale separation can be found. To be as explicit as possible, we restrict to vacua that can be constructed at leading order in $\alpha^\prime$ or in the 11D Planck length, i.e., we consider solutions of 10D supergravity with orientifold sources and solutions of 11D supergravity.

Whether string theory admits scale-separated AdS vacua in this setting is still under debate. 
There is a common belief that scale separation can only be achieved in the presence of orientifold sources, see \cite{Tsimpis:2012tu, Gautason:2015tig, Lust:2020npd, DeLuca:2021mcj} and references therein. This belief is based on lacking examples of scale-separated vacua without orientifolds.  Still, it can be explained if some assumptions about compactification geometries are made \cite{Gautason:2015tig}, see however \cite{DeLuca:2021mcj}. Furthermore, a number of swampland conjectures about scale separation in AdS have been proposed. In particular, the strong AdS distance conjecture of \cite{Lust:2019zwm} states that, for supersymmetric AdS vacua, there cannot be a parametric scale separation between the AdS curvature scale and the mass scale of an infinite tower of states. Further refinements of this conjecture were proposed that take into account log-corrections \cite{Blumenhagen:2019vgj} or the presence of discrete symmetries \cite{Buratti:2020kda}.

On the other hand, several constructions in the literature suggest that string theory does admit AdS vacua with a parametric scale separation. In 2005, DeWolfe, Giryavets, Kachru and Taylor (DGKT) constructed an infinite class of AdS vacua with some appealing properties in that regard \cite{DeWolfe:2005uu}.\footnote{These solutions were inspired by the earlier works \cite{Derendinger:2004jn, Behrndt:2004mj}.} The DGKT vacua were constructed in massive IIA supergravity compactified on Calabi-Yau (CY) 3-folds with intersecting O6 planes, RR $p$-form ($F_p$) and NSNS 3-form ($H_3$) fluxes. An interesting feature of these vacua is that the $F_4$ form flux quanta are unbounded. When these fluxes are sent to infinity, we find arbitrarily weak coupling, large volume and small cosmological constant, while the ratio between the KK scale $L_\text{KK}$ and the AdS length scale $L_H$ goes to zero. This suggests that these vacua are in very good control and achieve arbitrarily good scale separation.

The results of DGKT have been generalised in several ways. First, while the DGKT vacua possess $\mathcal{N}=1$ supersymmetry, there are also non-supersymmetric solutions with the same properties regarding control and scale separation \cite{Camara:2005dc, Narayan:2010em, Marchesano:2019hfb}. Similar results were furthermore found for compactifications of massive IIA on G$_2$ holonomy spaces down to three spacetime dimensions \cite{Farakos:2020phe} (see also \cite{Emelin:2021gzx}). 
Moreover, the DGKT solutions are part of a larger class of vacua for which the CY condition is relaxed to an SU$(3)$-structure condition \cite{Caviezel:2008ik}. Whether these vacua are scale-separated is case-dependent and can be rather subtle \cite{Font:2019uva} (see also \cite{Ishiguro:2021csu} for a recent related study). Finally, scale-separated AdS solutions are claimed to exist in IIB supergravity on SU$(2)$-structure geometries with intersecting O5/O7 planes \cite{Caviezel:2009tu, Petrini:2013ika}.

The focus of our paper is to revise and scrutinise some of the results regarding these non-CY solutions in IIA/IIB supergravity.
Concerning the IIB solutions, we rectify some statements made in the earlier works \cite{Caviezel:2009tu, Petrini:2013ika}. In particular, we explain that the existing solutions have vanishing cycles in the proper weak-coupling limit and hence provide no good backgrounds.

However, the main purpose of this paper is to strengthen the evidence for scale-separated AdS$_4$ vacua of IIA and M-theory. In particular, we construct a class of such solutions without Romans mass by performing two (formal) T-dualities of the DGKT solutions and their non-supersymmetric cousins. While this class of T-dual solutions was discovered before in \cite{Caviezel:2008ik}, we identify specific scaling limits where the backreaction of the O6 planes becomes arbitrarily small. This allows us to explicitly compute the backreaction at first order in a large-flux expansion using the techniques of \cite{Junghans:2020acz}. The class of solutions we study contains both weakly and strongly coupled limits. In the latter case, we also explicitly compute the lift to M-theory.

Our solutions are the first examples of scale-separated AdS solutions which neither require Romans mass nor a smearing of the O-planes. Thus, they address some worries about backreaction that have been expressed in the past \cite{Banks:2006hg, McOrist:2012yc}. 

Indeed, all previously known scale-separated AdS vacua were derived in the approximation where the O-plane backreaction is ignored. On general grounds, one constructs a superpotential and K\"ahler potential in the lower-dimensional supergravity and finds vacua of the resulting $F$-term potential. To obtain a consistent lift to 10-dimensional solutions, one needs to smear out the O-planes over the compact dimensions \cite{Acharya:2006ne, Grana:2006kf}. However, since O-planes live at the fixed loci of involutions, they cannot be smeared and the approximation seems bad. Nonetheless, such a smearing procedure is the standard way to ignore variations inside the internal dimensions, which are features at or above the KK scale and irrelevant to the 4D EFT.

One can worry that somehow the backreaction is worse than naively expected as claimed in \cite{Banks:2006hg, McOrist:2012yc}. However, it has recently been understood that there are regimes in which the backreaction is well-controlled and explicitly computable even for solutions with intersecting sources. In particular, \cite{Junghans:2020acz} and \cite{Marchesano:2020qvg}\footnote{This paper relied on earlier insights of \cite{Saracco:2012wc}.} computed the backreaction of O6 sources in the DGKT vacua at leading order in perturbation theory and found no obstacle to the existence of a solution. Analogous conclusions can be drawn in no-scale Minkowski compactifications \cite{Baines:2020dmu} (see also \cite{Blaback:2010sj}).

Nevertheless, one might wonder whether subtleties arise due to the presence of Romans mass. Indeed, the lack of an uplift to eleven dimensions hinders a non-perturbative understanding in that case. For that reason, we consider the non-CY AdS$_4$ solutions T-dual to DGKT since they exist without Romans mass. The class of solutions we construct can be argued to be lifted to 11D without the need of explicit sources. This seems to provide a counter-example to the assumptions made in the no-go theorem of \cite{Gautason:2015tig}. In particular, our solutions provide evidence for geometries which, despite having no separation of scales between internal and external curvature, do allow for a separation between AdS and KK scale.
Furthermore, our solutions are potential counter-examples to the strong AdS distance conjecture of \cite{Lust:2019zwm}.

This paper is organised as follows. In Section \ref{sec:scalesep10vs11D}, we explain the definition of scale separation and briefly review related no-go theorems and swampland conjectures. In Section \ref{scalsep}, we T-dualise the DGKT vacua to massless IIA and identify weakly and strongly coupled scaling limits with parametrically small backreaction. We also comment on earlier claims in the literature about scale separation in IIB. In Section \ref{sec:locIIA}, we compute the first-order backreaction of the IIA solutions, and in Section \ref{lift}, we lift the strongly coupled solutions to M-theory. In Section \ref{sec:purespinorloc}, we briefly discuss an alternative localisation method using the pure-spinor equations.  We conclude in Section \ref{concl} with a discussion of our results and future research directions. The details of several longer computations can be found in App.~\ref{examples}-\ref{App:GreensFunction}.
\\

\section{Scale separation in 10D and 11D}
\label{sec:scalesep10vs11D}
\subsection{Definition of scales and their separation}

Below, we recall some basic facts about the definition of scale-separated vacua and choices of units. We consider a general compactification from ten down to $D$ dimensions. It is useful to study the role of the string coupling and the volume. For that, we start with the 10D metric in string frame:
\be\label{planck}
\d s^2_{10} = \t_0^2 \t^{-2} \d s_D^2 + \r \d s^2_{10-D} \, .  
\ee
Here, $\rho^{(10-D)/2}$ denotes the volume in 10D string frame, which assumes that the metric $\d s^2_{10-D}$ describes a compact space with unit volume. Here and in the following, we will set $2\pi\ell_s=1$. The KK scale $L_\text{KK}$ is thus estimated to be
\be
L_\text{KK} = \sqrt{\rho}. \label{defkk}
\ee
The scalar $\t$ is defined as 
\be
\label{dilatontr}
\t^{D-2} = 2\pi\,{\rm e}^{-2 \phi} \r^{\frac{10-D}{2}}
\ee
such that $\d s^2_{D}$ is the metric in $D$-dimensional Einstein frame. For simplicity, we assume negligible effects from the warp factor and dilaton variations inside the extra dimensions, throughout this whole section. We will however be more careful about such effects in later sections. In our notation, $\tau_0$, $\rho_0$ describe the vacuum expectation values. In the literature on compactifications, one occasionally leaves out the compensating $\t_0^{2}$-factor in the reduction Ansatz. This then implies the choice of $D$-dimensional Planck units, as  can be verified from dimensional reduction: \be
S_D \supset \int_D\d^D x\sqrt{g_D}\left(\t_0^{D-2}\mathcal{R}_D + \ldots - \t_0^{D}V\right)\,,
\ee
where $\ldots$ represent all omissions, such as kinetic terms for the scalars, and $V$ is a (dimensionless) scalar potential. We conclude that the Planck scale is fixed by $M_p = \t_0$. In the vacuum, the Einstein equations tell us that
\be
\mathcal{R}_D = \frac{D}{D-2}\, M_p^2 V\,.
\ee
So, there are two length scales associated with this vacuum: the curvature radius $L_{H}$ and the ``inverse vacuum energy length scale'' $L_{\rho}$ defined as follows: \be
L_{H}^{-1} = \frac{M_p |V|^{1/2}}{\sqrt{(D-1)(D-2)}}\,,\qquad L_{\rho}^{-1}= M_p |V|^{1/D} \,.
\ee
Scale separation is then the condition that the KK scale decouples from the length scales of the non-compact manifold (either $L_{H}$ or $L_{\rho}$). In this paper, we are interested in a separation between $L_\text{KK}$ and $L_{H}$. This requires (in the vacuum):
\be \label{scalesep}
\frac{L_\text{KK}^2}{L_{H}^2} \sim \rho_0\tau_0^2 |V|\ll 1\,.
\ee
From here on, we assume full geometric moduli stabilisation and negative vacuum energy since the very existence of de Sitter solutions in string theory remains a substantially debated issue.\footnote{If we regard Minkowski vacua as formally having $L_H \rightarrow \infty$, then they are automatically scale-separated. Minkowski solutions with geometric moduli stabilisation are not known, and, without SUSY, it is unlikely they exist, see \cite{Palti:2020qlc}. With SUSY, all known vacua feature massless modes. The equivalence principle would hence be violated in such a lower-dimensional vacuum and lower-dimensional general relativity would not be recovered.}

If the criterion \eqref{scalesep} is satisfied, we can call a string vacuum genuinely lower-dimensional since then the KK masses are heavy compared to the AdS scale.
In particular, \eqref{scalesep} guarantees the existence of an energy range between the AdS scale and the KK scale where we have a consistent description in terms of a 4D effective field theory (EFT) involving a finite number of degrees of freedom. Note that, besides the KK tower, there may in general be further light states which could also invalidate the 4D EFT description of a given compactification. We will not study this possibility in this paper.

Of course, equation (\ref{scalesep}) is not sufficient for a decoupling of KK modes since it uses the overall volume as a proxy for the KK scale. Manifolds can be highly anisotropic when they have small and large cycles at the same time. We regard (\ref{scalesep}) as a necessary condition instead. In Section \ref{scalsep}, we will be more careful and consider the length scales associated to individual cycles.
Note also that our universe satisfies the generically much stronger constraint of scale separation between the cosmological-constant scale and the KK scale: $L_\text{KK}/L_\rho \ll 1$. This condition will not be satisfied in any of the models we discuss here.

\subsection{11D vs 10D}
\label{sec:11d10d}

Supergravity solutions with large string coupling are not to be trusted since stringy corrections are out of control. However, in IIA supergravity we can give meaning to a strongly coupled supergravity solution by uplifting it to 11D supergravity. The classical 11D solution can be trusted if curvatures are small. Below, we will uplift such strongly coupled solutions, so let us compare the definition of scale separation in the different duality frames. For simplicity, we will specialise to $D=4$.

Going from 11D to 10D \emph{string} frame proceeds via the Ansatz:
\be\label{uplifteq}
\d s^2_{11} = \e^{-\frac23 \phi}\d s^2_{10} + \e^{\frac{4}{3}\phi}(\d z +C_1)^2\,,
\ee
with $C_1$ the KK one-form that becomes the RR one-form of type IIA. The radion $\e^{\phi}$ is identified with the string coupling. The 7D volume is independent of $C_1$ and equal to $\rho^3\e^{-\frac43 \phi}$.\footnote{Our convention for the length of the M-theory circle is such that $2\pi\ell_p=2\pi\ell_s=1$.}

A necessary condition for scale separation in the M-theory frame (at strong IIA coupling) \emph{using 10D language} is given by the requirement:
\be\label{11Dscale}
\frac{\hat L^2_\text{KK}}{\hat L^{2}_{H}} = \frac{\rho^{\frac67}\e^{\frac27 \phi} }{L^{2}_{H}} \ll 1,
\ee
where $\hat{L}_H$ and $\hat{L}_\text{KK}$ are the M-theory AdS length scale (with $\hat L_H=\e^{-\frac13\phi}L_H$) and the volume scale of the 7D internal manifold, respectively.

At weak coupling, in the IIA duality frame, we instead require (\ref{scalesep}): $\rho L^{-2}_{H}\ll 1$. Because we look at manifolds with anisotropic scalings, the condition \eqref{11Dscale} is necessary but not sufficient. We give the precise requirement later.
However, the point of \eqref{11Dscale} is that the condition for scale separation at strong coupling can be stronger than the naive 10D condition \eqref{scalesep}, in particular when the M-theory circle is sufficiently large compared to $\rho$.

\subsection{A no-go theorem and its assumptions}
\label{sec:nogo}

It is well known that constructing Minkowski and de Sitter solutions with fluxes on a compact and static manifold requires at least orientifold singularities (sources) \cite{Maldacena:2000mw}. On the other hand, AdS vacua without orientifolds are easily obtained, but all such examples we know of are not scale-separated. Given that Minkowski vacua can be seen as a limit of scale-separated AdS vacua (since $L_\text{KK}$ remains finite whereas $L_H\rightarrow \infty$), one can wonder whether also orientifolds are required for scale-separated AdS vacua. One easily shows that 10D (and 11D) supergravity compactifications without orientifolds obey \cite{Gautason:2015tig} 
\be\label{nogo}
\left|\frac{\int \d^d y \sqrt{g_d}\, R_d}{\int \d^d y \sqrt{g_d}\, R_4}\right| < c \,, \qquad  R_d >0\,,
\ee
where the integral is over $d=6$ or $7$ compact dimensions, $R_d$ denotes the Ricci scalar of the internal manifold and $R_4$ the Ricci scalar of the external 4D manifold (in 10D string frame).
The number $c$ is always of order one. The result \eqref{nogo} implies that there is no separation of scales between the curvatures. In turn, this forbids scale separation as defined earlier in (\ref{scalesep}), \eqref{11Dscale} if one assumes that the internal curvature length scale $L_R$, defined as
\be
L_R^{-2} = \text{vol}_d^{-1}\int \d^d y \sqrt{g_d}\, R_d\,, \label{assump}
\ee
cannot be decoupled from the KK scale (i.e., from $L_\text{KK}$ for $d=6$ and $\hat L_\text{KK}$ for $d=7$). Here we again define the KK scale as the volume scale (cf.~\eqref{defkk}, \eqref{11Dscale}), which, as explained above, can of course differ from the actual KK scale. Recently, a much more in-depth and precise analysis appeared in \cite{DeLuca:2021mcj} that took into account warping effects and found them to be possibly relevant.  

It naively seems easy to decouple the curvature scale from the volume scale by simply considering a Ricci-flat compact space. However, the supergravity equations insist on a positive Ricci tensor. To obtain scale separation without orientifolds, we want to evade the aforementioned assumption \eqref{assump}. We thus require a family of compact manifolds with positive Ricci tensor for which the ratio $L_\text{KK}/L_R$ can be made parametrically small (not large). In other words, we want to shrink the KK scale at fixed curvature. We are not aware of a single smooth Riemannian manifold with this property, while at the same time the Ricci tensor is positive definite. For pedagogical purposes, Appendix \ref{examples} contains five classes of simple example metrics where the ratio $L_\text{KK}/L_{R}$ is studied. Only certain examples with non-positive-definite Ricci tensors were found to allow a decoupling in the right direction.

To our knowledge, no examples of flux geometries that bypass the assumptions of the no-go theorem in \cite{Gautason:2015tig} for scale-separated flux vacua are known in the literature. However, this does not mean that no counterexamples could exist. In fact, an important result of this paper is a concrete proposal for such a counterexample, which is obtained by lifting orientifold flux vacua of 10D IIA supergravity to 11D supergravity. The resulting 7D internal geometries will circumvent the assumptions in the no-go theorem, i.e.,~the curvature and KK scale will be decoupled, but at the expense of a singular geometry. The singularity is not supported by an explicit source in 11D. As we will explain in more detail below, it is natural to assume that it is resolved in M-theory in the same way as the singularities of O6 planes in flat space. However, we remain open-minded to the possibility that this could be false.

\subsection{Swampland considerations}

Guided by a generalisation of the distance conjecture \cite{Ooguri:2006in} and by explicit examples in string theory, it was argued in \cite{Lust:2019zwm} that AdS vacua in string theory come with a tower of states whose mass scale $m$ behaves as:
\be\label{AdSconj}
\Lambda\rightarrow 0\quad \longrightarrow\quad m\sim |\Lambda|^{\alpha}\qquad \text{in Planck units}\,, 
\ee
with $\alpha>0$ and $\Lambda$ being the value of the scalar potential $V$ in the AdS minimum.\footnote{Reference \cite{Blumenhagen:2019vgj} argued for certain ``log corrections" to the conjectures.}
In particular, it is conjectured that $\alpha=1/2$ for supersymmetric AdS vacua. The latter case, which is also referred to as the \emph{strong} AdS distance conjecture, would not allow for a separation of scales between the Hubble length and the tower with mass scale $m$ since the product $m\,L_H$ remains of order one. If correct, the DGKT vacua are in the swampland and something in their construction would have to be flawed.\footnote{Alternatively, a tower other than the KK tower restores the strong ADC, but no obvious candidate tower has been identified. In principle, it could also be possible that the SUSY DGKT solutions break supersymmetry when the O6 planes are fully localised.}

Note that $m$ in equation (\ref{AdSconj}) is not necessarily the KK scale but more generally refers to the mass scale of any light tower of states. Indeed, as stated before, the phenomenological condition of scale separation for AdS vacua is aimed at being able to write down an EFT with finite degrees of freedom. In this paper, we focus on the specific case in which the tower of states is the KK tower. 

Reference \cite{Buratti:2020kda} suggested an intriguing refinement of \eqref{AdSconj} that allows scale separation under the condition that discrete gauge symmetries are present. The refined conjecture is then consistent with the DGKT vacua but inconsistent with the 3D AdS vacua of \cite{Farakos:2020phe} that are constructed from G$_2$ orientifolds \cite{progress}. Below, we will discuss solutions that are T-dual to the original DGKT solution and lift them to M-theory. The discrete gauge symmetries of the DGKT solution are invariant under string dualities and therefore one can show that our solutions satisfy the refined conjecture of \cite{Buratti:2020kda} but violate equation \eqref{AdSconj} for $\alpha=1/2$.

Finally, a related but weaker conjecture appeared in \cite{Gautason:2018gln}. It states that AdS vacua always come at least with one field $\phi$ that is light in the sense that the mass $m_{\phi}$ cannot be parametrically heavy in AdS units:
\be \label{AdSconj2}
m^2_{\phi}L_{H}^2 \approx \mathcal{O}(1)\,.
\ee
As explained in \cite{Gautason:2018gln}, this has relevance for supersymmetry breaking and might explain the difficulties in finding controlled dS vacua from uplifting mechanisms. This weaker conjecture (\ref{AdSconj2}), named the AdS/moduli conjecture in \cite{Blumenhagen:2019vgj}, is furthermore closely related to the refined dS conjecture \cite{Andriot:2018wzk, Garg:2018reu, Ooguri:2018wrx} when applied to negative potentials. Note that DGKT vacua are consistent with (\ref{AdSconj2}).

\section{Scale-separated solutions from smeared orientifolds}
\label{scalsep}
\subsection{Scale separation in IIA}\label{sec:weak}

Historically, the first flux vacua with scale separation were found in the context of Romans IIA supergravity compactified on (generalised) Calabi-Yau spaces with O6 planes. This setup was analysed in \cite{Derendinger:2004jn, Villadoro:2005cu, Behrndt:2004mj} and then more extensively in \cite{DeWolfe:2005uu}, where it was realised that the solution becomes scale-separated in the weak-coupling limit. Note that in these AdS vacua all geometric moduli are stabilised. They are hence quite remarkable in many aspects. A 10D understanding of these solutions turns out to be much simpler and reveals that the solutions are really only understood in the limit in which the O6 planes can be regarded as smeared \cite{Grana:2006kf, Caviezel:2008ik} (see also \cite{Acharya:2006ne}). It is often considered to be a problematic feature when O-planes are smeared, but this is potentially based on certain misconceptions. Certainly, O-planes are localised objects in string theory and, unlike D-branes, cannot be stacked. However, the smearing is only a formal device to find a 1-1 map between the lower-dimensional supergravity and the 10D equations of motion. The smearing approximation amounts to solving the integrated 10D equations and is expected to approximate the true solution in the weak-coupling, large-volume limit. Indeed, this expectation was explicitly confirmed in \cite{Junghans:2020acz, Marchesano:2020qvg}, where
`first-order' backreacted solutions were found for the DGKT vacua, demonstrating how the smeared solution becomes better approximated in the wanted limit of weak coupling. The same behaviour was also verified in \cite{Baines:2020dmu} for no-scale Minkowski vacua in massive IIA supergravity. 

In what follows, we briefly review a subclass of solutions of IIA supergravity with fluxes and O6 planes, defined by the condition that the internal manifold has an SU$(3)$-structure group. We will recall the general solution from a 10D viewpoint, as found in \cite{Grana:2006kf}.  Many more details, such as the 4D moduli stabilisation, can be found in several references, such as \cite{DeWolfe:2005uu, Camara:2005dc,  Caviezel:2008ik}.

The general form of the 10D supersymmetric solutions was found in \cite{Lust:2004ig, Grana:2006kf} (see also \cite{Behrndt:2004mj} for earlier work). It involves NSNS 3-form flux $H_3$ and the RR fluxes $F_0,F_2, F_4, F_6$ and can be written in terms of the K\"ahler 2-form $J$,  the complex-structure 3-form $\Omega$ and the torsion classes $\mathcal{W}_i$ (in string frame):\footnote{Our normalisations of $J$ and $\Omega$ are such that $\star_6 J = - \frac{1}{2} J \wedge J$, $\star_6 \Omega = - \rmi \Omega$, $\Omega \wedge \bar{\Omega} =- \frac{4}{3}\rmi J^3 = 8 \rmi~\text{dvol}_6$, and we set $g_s=\e^{\phi}$.}
\begin{align}
\label{eq:H}
H_3 &= 2m   \, \Re \Omega,\\
\label{eq:F0}
g_s F_0 &= 5 m,\\
\label{eq:F2}
g_s F_2 &= \frac{\tilde{m}}{3} J + \rmi \mathcal{W}_2,\\
\label{eq:F4}
g_s F_4 &= \frac{3}{2} m J \wedge J,\\
\label{eq:F6}
g_s F_6 &=  3\tilde{m} \text{ dvol}_6.
\end{align}
We see that there are two parameters $m$ and $\tilde{m}$ that also fix the 4D AdS curvature and part of the torsion:
\begin{align}
\label{eq:dJ}
 \dd J &= 2 \tilde{m} \, \Re \Omega\,,\\
\label{eq:dOmega}
\d \Omega &= -\frac{4}{3} {\rm i}\: \tilde{m} J \wedge J + \mathcal{W}_2 \wedge J\,,\\
 \frac{1}{L_{H}^2} &= m^2 + \tilde{m}^2\,.
\end{align}
The source term describing the O6 planes (and possibly D6 branes) in the $F_2$ Bianchi identity (and also in the Einstein equations and the dilaton equation) is given by:\footnote{This is the source that enters the equation $\d F_2=F_0H_3+j_3$. It is related to the conventions of Section \ref{sec:locIIA} and Appendix \ref{appEOM} as follows: $j_3=-2\sum_i j_{i3}$, $\text{Im}\Omega = \sum_i \text{dvol}_{\pi_i}$ and $g_s \star_6 (\text{Im}\Omega \w j_3)= -2g_s \sum_i j_{\pi_i}=32q$, where
$q$ is the O6-charge density.}
\be
g_sj_3 = {\rm i}\,\d\mathcal{W}_2 +\left(\frac 23 \tilde m^2 - 10 m^2\right)\Re \Omega. \label{j3}
\ee
It can be shown that this form corresponds to the smearing of a source that captures 4 mutually intersecting (sets of) O6 planes \cite{Caviezel:2008ik} (consistent with orientifold involutions and projections). One can easily derive that \cite{Tsimpis:2012tu, Petrini:2013ika}:
\begin{equation}
\frac{R_6}{R_4} =-\frac{1}{6} \frac{6\tilde{m}^2+10 m^2 - 8|q|}{m^2+\tilde{m}^2},
\end{equation}
where $q$ is the O6-charge density. So, clearly without O6 planes we cannot achieve scale separation of the curvatures.  

The solutions on Calabi-Yau spaces fall into the class with $\tilde{m}=0$ and $\mathcal{W}_2=0$. Let us recall how scale separation is achieved there.  The Bianchi identities leave the $F_4$ flux unconstrained, whereas $F_0$ is bounded by the RR tadpole \cite{DeWolfe:2005uu}. Let us therefore introduce a (discrete) parameter $n$ such that
\be
F_4 \sim n, \qquad F_0 \sim n^0.
\ee
One can check from \eqref{eq:F0}, \eqref{eq:F4} that the other quantities must scale like
\be
g_s, \: m \sim n^\alpha\,, \qquad J\sim n^{1/2}\,,\qquad \text{dvol}_6\sim  n^{3/2} \,.
\ee
We can fix $\alpha$ from the source form \eqref{j3} since we cannot let the O6 charges scale. This gives $\Re \Omega \sim n^{-\alpha}$. Assuming $\Im \Omega \sim \Re \Omega \sim n^{-\alpha}$, we then find from $J^3 \sim \Omega \w \bar\Omega$  that $\alpha = -3/4$. We have thus shown that supersymmetric solutions of the above type come in families with a free parameter $n$. The large-$n$ limit amounts to weak coupling, large volume \emph{and} scale separation since
\be
\frac{L^2_\text{KK}}{L^2_H} \sim n^{-1} \rightarrow 0.
\ee
One can show that the smeared IIA equations of motion (cf.~Section \ref{sec:locIIA} and Appendix \ref{appEOM}) are invariant under the $n$ rescalings even without imposing supersymmetry. The same large-$n$ limit can therefore also be taken in non-supersymmetric solutions \cite{Marchesano:2019hfb}.

Clearly, the Calabi-Yau solutions are only a subset out of many possibilities. In particular, putting the torsions ($\tilde{m}$, $\mathcal{W}_2$) to zero eliminates $F_2$ and $F_6$.\footnote{Note that $F_2$ and $F_6$ denote the \emph{improved} field strengths here. On the other hand, the associated flux numbers in the flux superpotential ($e_0$ and $m_a$ in the notation of \cite{DeWolfe:2005uu}) are non-zero in general.} When the torsions are non-zero, we move away from the CY limit and the situation is more complicated. In particular, the O6/D6 sources are still (generalised) calibrated \cite{Koerber:2007jb} but they do not have to wrap cycles that are non-trivial in homology over the real numbers. For instance, there are some simple examples in which they wrap cycles that are non-trivial in homology over the integers (torsional cycles) \cite{Marchesano:2006ns}. This makes it consistent to have solutions with zero Romans mass despite the naive RR tadpole coming from the $F_2$ Bianchi identity. Since we are interested in solutions that can be lifted to 11D supergravity, we pay particular attention to the set that has $F_0=0$ and thus $m=0$. A particularly interesting example of this family can be found by a formal double T-duality, as was suggested in \cite{Banks:2006hg} and made very explicit and simple in \cite{Caviezel:2008ik}. When the original solution has a toroidal covering space, the double T-dual solution is an orbifold of a nilmanifold.

Let us go through some of the steps and start off with the flat metric on $T^6$:
\be
	\dd s_6^2 = (L_1e^1)^2 + (L_2 e^2)^2 + (L_3 e^3)^2 + (L_2 e^4)^2 + (L_3 e^5)^2 + ( L_1 e^6)^2,
\ee
where $e^m =\dd y^m$.  For simplicity of the presentation, we only keep 3 deformation parameters $L_1$, $L_2$ and $L_3$. 

We choose the O6 involution such that we have 4 sets of O6 planes\footnote{More precisely, there are 8 parallel O-planes in each direction and thus 32 O-planes in total (see also Appendix \ref{App:GreensFunction}).} wrapping the cycles defined by the following volume forms:\footnote{The actions of the orientifold involution and the orbifolding on the $e^m$ in our conventions are stated, e.g., in \cite{Danielsson:2011au} in Subsection 4.2.}
\be
\label{eq:O6volumeforms}
e^1 \wedge e^2 \wedge e^3 \,,\qquad e^3 \wedge e^4 \wedge e^6\,,\qquad e^2 \wedge e^5 \wedge e^6\,,\qquad e^1 \wedge e^4 \wedge e^5 \,.
\ee
We can then use the formulas in \cite{Danielsson:2011au} to find the following SU(3)-invariant forms:
\begin{align}
	J &= -L_1^2 e^{16} - L_2^2 e^{24} + L_3^2 e^{35}\,,\\
	\Re \Omega &=  L_1 L_2 L_3 \left( e^{456} + e^{236} - e^{134} - e^{125}\right)\,,\\
	\Im \Omega &=   L_1 L_2 L_3 \left(e^{123} + e^{145} + e^{256} + e^{346} \right)\,,
\end{align}
where, e.g., $e^{16}\equiv e^1 \wedge e^6$ and similarly for the other forms.
The fluxes are given as in \eqref{eq:H}-\eqref{eq:F6} with $\tilde{m} = 0$ and $\mathcal{W}_2=0$.

If we T-dualise along directions 1 and 6, we obtain the metric
 \be
	\dd s_6^2 = (L_1^{-1} e^1)^2 + (L_2 e^2)^2 + (L_3 e^3)^2 + (L_2 e^4)^2 + (L_3 e^5)^2 + ( L_1^{-1} e^6)^2,
\ee
where now $e^1$ and $e^6$ are not closed:
\begin{equation}
\label{geomflux}
\dd e^{1} = -e^{23} - e^{45}\,,\qquad \dd e^{6} = - e^{34} - e^{25}.
\end{equation}
Note that we relabeled $1 \leftrightarrow 6$ and set the structure constants to $-1$.\footnote{The two T-dualities also change the cycles that the O6 planes wrap but, after relabelling the directions 1 and 6 via $1 \leftrightarrow 6$, the O6 planes have the same volume forms as in the original expression \eqref{eq:O6volumeforms}. Depending on which flux numbers one chooses in the DGKT solution, the structure constants $f^1{}_{23}=f^1{}_{45}=f^6{}_{34}=f^6{}_{25}\equiv f$ on the T-dual side can differ. We fix $f=-1$ here, which can be done by an appropriate rescaling of the vielbeins (see also Appendix \ref{App:GreensFunction}).}
The SU(3)-structure forms become
\begin{align}
\label{J}
	J &= -L_T^{2} e^{16} - L_2^2e^{24} + L_3^2 e^{35}\,,\\
	\label{OmegaR}
	\Re \Omega &=  L_T L_2 L_3 \left( e^{456} + e^{236} - e^{134} - e^{125}\right)\,,\\
	\label{OmegaI}
	\Im \Omega &=   L_T L_2 L_3 \left(e^{123} + e^{145} + e^{256} + e^{346} \right)\,,\\
	 \mathcal{W}_2 &= \frac{8 \tilde{m}}{3}\rmi \left(-2L_T^{2}e^{16} +L_2^2e^{24}- L_3^2 e^{35}\right)\,, \label{w2}
\end{align}
where we defined $L_T \equiv L_1^{-1}$. The fluxes are as in \eqref{eq:H}-\eqref{eq:F6}, with $m = 0$ and $\tilde m =\frac12 L_T/(L_2 L_3)$. The $F_2$ Bianchi identity reduces to
\begin{equation}
\label{eq:BISU3}
g_s j_3 + 10 \tilde m^2 \Re \Omega =0.
\end{equation}
The above geometry is known as an Iwasawa manifold and was studied in \cite{Caviezel:2008ik} (see also \cite{Derendinger:2004jn, Villadoro:2005cu, Camara:2005dc}). 

Using \eqref{eq:dJ}-\eqref{j3}, one infers that this solution is part of a larger class of solutions with $m=0$, $\tilde{m}\neq 0$ with the following scaling behaviour:
\begin{align}
\label{eq:general_scalings}
\begin{split}
    &L_T \sim n^{(a-b-c)/4}, \qquad L_2 \sim n^{(a-b+c)/4}, \qquad L_3 \sim n^{(a+b-c)/4}\,,\\
    & L_H \sim n^{(a+b+c)/4}, \qquad g_s \sim n^{(a -3b - 3c)/4},
\end{split}
\end{align}
where the scaling exponents $a$, $b$ and $c$ are free parameters.
These scalings are engineered in exactly such a way that the current $j_3$ does not scale and hence the orientifold charge stays fixed at order one.
Using the above expressions in \eqref{eq:H}-\eqref{eq:F6}, one finds that the RR form fields scale as follows: $F_6$ scales like $n^a$, the component of $F_2$ along $e^{35}$ scales as $n^b$, the component along $e^{24}$ scales as $n^c$ and the component along $e^{16}$ does not scale.
We therefore require $a,b,c\ge 0$ so that the large-$n$ limit is compatible with flux quantisation. In what follows, we take without loss of generality $c \geq b$. We will furthermore assume $b>0$. As we will see below, this is required to control the backreaction.

Our choice of scaling exponents is such that the second two-torus, with associated length scale $L_2$, is the largest one (if $c=b$, the third two-torus, with length scale $L_3$, is equally large).
We will therefore assume that the KK scale is given by $L_\text{KK} \sim L_2$. One may wonder whether the presence of the geometric fluxes invalidates this naive estimate of the KK scale. This can be checked by considering the spectrum of the Laplacian on our twisted torus. To this end, we generalised the 3D formulae of \cite{Andriot:2016rdd} to our 6D case with several geometric fluxes (see Appendix \ref{appKK}). The result is that the geometric fluxes increase the eigenvalues of some of the KK states by terms of the order $1/L_2L_3$. However, the lowest-lying states still have eigenvalues $\sim 1/L_2^2$ as on a flat torus so that our  estimate $L_\text{KK} \sim L_2$ is correct.

As reviewed in Section \ref{sec:scalesep10vs11D}, scale separation requires the AdS scale $L_H$ to grow faster than the KK scale $L_\text{KK}$ in the large-$n$ limit. We find
\be
    \frac{L^2_\text{KK}}{L^2_H} \sim n^{-b}.
\ee
Since $b>0$, the solution is scale-separated for every choice of $a$, $b$ and $c$.
Note that scale separation with respect to the KK scales of the other two-tori (with length scales $L_T$ and $L_3$) is automatically implied here since $L_2$ is the largest length scale.

As can be seen from \eqref{eq:general_scalings}, weakly coupled solutions exist when $b+c > \frac{a}{3}$. As an example, we take $L\equiv L_2=L_3$ such that $b=c$. When $a = 1,~b=c=\frac{1}{4}$, we find: 
\be
     L_T \sim n^{\frac{1}{8}}, \qquad L \sim n^{\frac{1}{4}}, 
     \qquad L_H \sim n^{\frac{3}{8}}, \qquad g_s\sim n^{-\frac{1}{8}}.
\ee
We also have large volume because $\vol_6 \sim L_T^2 L^4 \sim n^{\frac{5}{4}}$. This shows that weak coupling, large volume and scale separation are  possible in \emph{massless} type IIA supergravity.\footnote{In \cite{Caviezel:2008ik}, a different scaling limit involving a rescaling of the O6 plane charges was argued to lead to scale-separated vacua. However, these charges are fixed in string theory so that we do not consider such limits here.}

There also exist scale-separated solutions at strong coupling suitable for an uplift to 11D. Because of \eqref{eq:general_scalings}, (parametrically) strong coupling implies $b+c < \frac{a}{3}$.  Consider for instance equation \eqref{eq:general_scalings} with $a = 1,~b = c = \frac{1}{8}$, then:
\be
L_T \sim n^{\frac{3}{16}}\,,\qquad L \sim n^{\frac{1}{4}}\,,\qquad L_H \sim n^{\frac{5}{16}}\,,\qquad g_s \sim n^{\frac{1}{16}}\,.
\ee
This corresponds to strong coupling, large volume and scale separation from a IIA viewpoint. However, such a viewpoint is not attainable anymore and this solution is properly described by a weakly curved solution in 11D. As we will see below, a consistent lift requires localising the orientifold planes. In particular, the backreaction is precisely such as to make $F_2$ closed, such that we can locally write $F_2=\d C_1$ and use the standard dictionary for the 11D metric \eqref{uplifteq}. We will also see that the localisation will not alter the values of the coupling or overall volume, up to parametrically small corrections. We can therefore already verify whether or not the solution passes the criterion of 11D scale separation discussed in Section \ref{sec:11d10d}. We find indeed that it does, because the length scale of the second torus, $g_s^{-1/3}L$, is larger than the one of the M-theory circle, $g_s^{2/3}$.
Because of \eqref{uplifteq}, the KK and AdS scales in the 11D and 10D metrics are then related by $\hat L_\text{KK} = g_s^{-1/3} L_\text{KK} \sim g_s^{-1/3} L$ and $\hat L_H = g_s^{-1/3}L_H$.
We thus have
\be\label{scalesep11D}
\frac{\hat L^2_\text{KK}}{\hat L^2_{H}}  \sim  n^{-\frac{1}{8}}\rightarrow 0\,.
\ee
Similarly to the 10D case, one may again be worried that the non-trivial fibration of the M-theory circle for $C_1\neq 0$ could affect our estimates of the KK scales.
In particular, the non-closure of $e^z=\d z+C_1$ is given by $F_2$, which has components scaling non-trivially with $n$ (see the discussion below \eqref{eq:general_scalings}) and might thus have a large effect on the KK spectrum. Applying the formulae of \cite{Andriot:2016rdd} to our 7D case, we find that $F_2$ increases the squared masses of the KK modes along the M-theory circle by subleading terms of the order $\tilde m/g_s^{1/3}\sim n^{-1/3}\ll g_s^{-4/3}\sim n^{-1/12}$. However, the lightest KK modes do not receive such corrections so that $\hat L_\text{KK}\sim g_s^{-1/3}L$ as on a flat torus.

There is one more crucial aspect that needs to be checked in order to consider these vacua as possibly meaningful. Since they are derived in the smeared approximation, one needs to make sure that smearing can be justified here, given that large coupling potentially implies significant  backreaction of the O6 sources. The smearing approximation relies on a limit in which the smeared solution approximates the localised solutions arbitrarily well, away from the singularities \cite{Baines:2020dmu,Marchesano:2020qvg,Junghans:2020acz}. Localised sources tend to deform their neighbouring spacetime up to a characteristic distance after which the deformation quickly dies out. For an O6 plane in flat space, this distance is $g_s$ in string units and it is thus small at weak coupling.\footnote{Here we mean the distance measured in the unbackreacted Minkowski metric. Note that the warped metric becomes ill-defined near the O-plane in the supergravity approximation.} In order to verify whether the backreaction is small within a compactification, one needs to compare this distance with the length scale of the cycle transverse to the orientifold plane. If we call the latter length scale $\bar{L}$, then we require $g_s/\bar{L}$ to be small. Indeed, it was explicitly verified for concrete examples in \cite{Marchesano:2020qvg, Junghans:2020acz, Baines:2020dmu} that the supergravity fields, symbolically denoted $S$, organise themselves in an expansion of the form
\be
S=S_0 + \sum_{i>0} c_i \left(\frac{g_s}{\bar{L}}\right)^i\,,
\ee
with $S_0$ the smeared solution. In DGKT, one has $g_s/\bar{L}\sim 1/n$ \cite{Marchesano:2020qvg,Junghans:2020acz}.

For each of the O6 planes in our setup, taking $\bar{L}$ to be the volume scale of the orthogonal spaces yields $\bar{L}=(L_TL_2L_3)^{1/3}$. We thus find
\be
\frac{g_s}{\bar{L}} \sim n^{-\frac{2(b+c)}{3}} \rightarrow 0  \,. \label{gsl}
\ee 
We stress that this is a necessary but not a sufficient condition. Indeed, we will see below that, in our solutions, the backreaction is somewhat stronger than the above estimate. The reason is that some of the cycles in our compactification space grow much faster than others at large $n$ such that the transverse space of the O6 planes is effectively less than three-dimensional in this limit. As will be shown in Section \ref{sec:locIIA} and Appendix \ref{App:GreensFunction}, the backreaction scales like
\begin{equation}
\frac{g_sL_2}{L_TL_3} \sim n^{-b}
\end{equation}
in our case, which is larger than \eqref{gsl} (recall that $c\ge b$) but remains small in the large-$n$ limit.

We conclude that, despite the strong coupling, the smeared approximation is at least as well motivated as for the weakly coupled solutions. This means that, as a pure supergravity solution, the backgrounds might make sense. Of course, those backgrounds are not meant to be trusted as IIA string theory solutions but rather as M-theory backgrounds. The rest of this paper is devoted to studying the localisation and uplift of these solutions. However, before doing so, we briefly recall and revise what is known about scale-separated vacua in IIB supergravity.

\subsection{Scale separation in IIB?}

Reference \cite{Petrini:2013ika} claims that SUSY scale-separated solutions of type IIB supergravity can be found from SU$(2)$-structure backgrounds with smeared O5 and O7 planes. Upon a closer look, we find no such solutions in the considered set-ups that  have a parametrically large volume  of \emph{all} cycles and at the same time parametrically weak coupling. Since reference \cite{Petrini:2013ika} focussed on the overall volume and not on all individual cycles, this was left unnoticed. Looking at all the different examples in the literature \cite{Caviezel:2009tu,Petrini:2013ika}, we find that there is no known example in which all individual volumes in the internal space become parametrically large while the string coupling becomes parametrically weak and we have parametric scale separation. Hence the existence of such solutions, i.e.,~the analogue of the DGKT \cite{DeWolfe:2005uu} solution in type IIB, remains an open problem.\footnote{See \cite{Emelin:2021gzx} for recent work on scale separation in three-dimensional flux vacua of IIB on co-calibrated G$_2$ orientifolds.}

\section{First-order localisation of IIA solutions}
\label{sec:locIIA}

The solutions presented in the previous section were obtained in the smeared approximation. In particular, this means that the dilaton and the warp factor $\e^{A}$ are constant.
When the sources are localised, we expect them to acquire a non-trivial profile along the internal manifold. It is thus convenient to restore the warp factor explicitly in the 10D string-frame metric:
\begin{equation} \label{IIAmetric}
    \d s^2_{10} = w^2 \tilde g_{\mu \nu}\d x^\mu \d x^\nu + g_{mn} e^m e^n, \qquad\qquad w(y) = L_H \e^{A(y)} ,
\end{equation}
where $\tilde g_{\mu\nu}$ is the unwarped AdS$_4$ metric with unit radius, while the indices $m,n=1,\dots 6$ run over the internal directions. The internal metric $g_{mn}(y)$ also differs from the smeared solution. We define the vielbeins as in Section \ref{sec:weak}.

In what follows, we describe the first-order localisation of the scale-separated AdS$_4$ solutions presented in Section \ref{sec:weak}, for the case $m=0$. This follows the same logic as the first-order localisation for the case $\tilde{m}=0$ carried out in \cite{Junghans:2020acz}. In order to be self-contained, we recall the bosonic IIA field equations with sources in Appendix \ref{appEOM}.
The most important equations will be the $F_2$ Bianchi identity and the Einstein and dilaton equations, as they contain the O6 sources. For later convenience, we state them here in the smeared approximation:
\begin{align}
0 & = \d F_2 + 2 \sum_i j_{i3}, \label{eq:smear0} \\
\label{eq:smear1}
    0 & = 12 \frac{\mathcal{T}^2}{w^2} - \sum_{q=2,6} \frac{q-1}{4}|F_q|^2+\frac12 \mathcal{T}\sum_i j_{\pi_i}, \\
    \label{eq:smear2}
    0 &=-\mathcal{T}^2 R_{mn} + \frac12 \sum_{q=2,6} \left(|F_q|^2_{mn} - \frac{q-1}{8}g_{mn}|F_q|^2\right) + \sum_i \left(\Pi_{i, \, mn} - \frac78 g_{mn}\right)\mathcal{T} j_{\pi_i},\\
    \label{eq:smear3}
    0&= -24 \frac{\mathcal{T}}{w^2} + 2 \mathcal{T}R_{mn}g^{mn} + 2 \sum_i j_{\pi_i},
\end{align}
where we set $\mathcal{T} \equiv \e^{-\phi}$ and used that $H_3=F_0=F_4=0$ in the smeared solution. Furthermore, $\Pi_{i,mn}$ is the projector onto the world-volume of the $i$-th source (with $\Pi_{i,mn}=g_{mn}$ for directions parallel to the source and zero otherwise). The $j_{\pi_i}$'s are constants which represent the smeared sources. In the localised equations, the sources are properly described by delta distributions. Our conventions for smeared and localised sources are stated in Appendix \ref{appEOM}.

The main strategy in this section will be to first make an Ansatz for an expansion describing the localisation of the sources order by order. Using this Ansatz, we will then show that all equations of motion reduce to a single Poisson equation, which can  be solved explicitly.

\subsection{Ansatz}
\label{sec:ansatz}

In the following, we denote by subscripts $1$, $2$, $3$ the three (twisted) 2-tori with volumes $L_T^2$, $L_2^2$ and $L_3^2$, respectively. For example, $g_{mn,1}$ denotes the 2D block of the metric corresponding to the first torus, i.e., $g_{1}= \text{diag}(g_{11},g_{66})$, etc. Let us also decompose $F_2$ according to which legs it has on each of the three 2-tori:
\begin{equation}
F_2 = F_{2,1}+ F_{2,2}+ F_{2,3} + F_{2,12}+ F_{2,23}+
F_{2,31}.
\end{equation}
For example, $F_{2,1}$ means both legs on the first torus, while $F_{2,12}$ means one leg on the first torus and one leg on the second one. The last three components in the above expression for $F_2$ are absent in the smeared approximation but are generated by backreaction.

Following \cite{Junghans:2020acz}, we can now make an Ansatz for the backreacted solution in the large-$n$ regime, at first order. We already know from Section \ref{sec:weak} how the fields in the smeared approximation scale with $n$. The non-trivial part is finding the scaling with $n$ of the first-order correction. However, we will see below that the scaling we choose is consistent with the equations of motion.\footnote{The general logic is that coefficients at order $m$ in perturbation theory are sourced by the fields at order $m-1$ and solved at order $m-1$ in the equations of motion.}

For the dilaton and warp factor, we have:
\begin{align}
\e^{-\phi} \equiv \mathcal{T} &= n^{(3b+3c-a)/4} \left[\mathcal{T}^{(0)} + \mathcal{T}^{(1)}n^{-b} + \mathcal{O}(n^{-2b})\right], \label{ansatzw} \\
 L_H \e^{A} \equiv w &= n^{(a+b+c)/4} \left[w^{(0)} + w^{(1)}n^{-b} + \mathcal{O}(n^{-2b})\right]\,. \label{sol}  
\end{align}
Here $\mathcal{T}^{(i)}$, $w^{(i)}$ are the expansion coefficients of our large-$n$ expansion, which themselves do not scale with $n$. In particular, $\mathcal{T}^{(0)}$ and $w^{(0)}$ denote the (constant) dilaton and warp factor in the smeared approximation, \emph{but with the $n$-scaling taken out}. $\mathcal{T}^{(1)}$ and $w^{(1)}$ are functions of the internal coordinates $y^m$ and encode the backreaction. This notation facilitates the order-by-order analysis since it is an expansion in an inverse power of $n$. 

Similarly, we write
for the metric components:
\begin{align}
g_{mn,1} &= n^{(a-b-c)/2} \left[g_{mn,1}^{(0)} + g_{mn,1}^{(1)}n^{-b} + \mathcal{O}(n^{-2b})\right], \\
g_{mn,2} &= n^{(a-b+c)/2} \left[g_{mn,2}^{(0)} + g_{mn,2}^{(1)}n^{-b} + \mathcal{O}(n^{-2b})\right], \\
g_{mn,3} &= n^{(a+b-c)/2} \left[g_{mn,3}^{(0)} + g_{mn,3}^{(1)}n^{-b} + \mathcal{O}(n^{-2b})\right], \label{ansatz}
\end{align}
where $g_{mn,1}^{(0)}=L_T^{(0)2}\delta_{mn}$ and analogous relations hold for the other components.
For $F_2$ and $F_6$, we write:
\begin{align}
F_6 &= n^a \left[F^{(0)}_6 + \mathcal{O}(n^{-b})\right], \label{ansatz0} \\
F_{2,1} &= n^{0}\left[F_{2,1}^{(0)} + \mathcal{O}(n^{-b})\right], \\
F_{2,2} &= n^{c}\left[F_{2,2}^{(0)} +F_{2,2}^{(1)}n^{-b}+ \mathcal{O}(n^{-2b})\right], \\
F_{2,3} &= n^{b}\left[F_{2,3}^{(0)}  +F_{2,3}^{(1)}n^{-b}+\mathcal{O}(n^{-2b})\right], \\
F_{2,12} &= n^{(c-b)/2}\left[F^{(1)}_{2,12} + \mathcal{O}(n^{-b/2})\right], \\
F_{2,23} &= n^{(c-b)/2}\left[ F^{(1)}_{2,23} + \mathcal{O}(n^{-b/2})\right], \\
F_{2,31} &= n^0\left[F^{(1)}_{2,31} + \mathcal{O}(n^{-b/2})\right]\,. \label{ansatzf}
\end{align}

The overall scalings with $n$ are fixed by what we found earlier for the smeared solution in Section \ref{sec:weak} (aside from $H_3$, $F_0$ and $F_4$ which are zero for $m=0$). We keep $F_0=0$ at all orders since it is a quantised constant which is not supposed to receive $1/n$ corrections. Furthermore, we have not written out the expansion coefficients of $F_4$ and $H_3$ since keeping them zero is consistent at the order we are working with. Our qualitative conclusions are insensitive to the actual values of these coefficients. For instance, in the M-theory lift, they would be sub-leading corrections in the expressions for the 11D 4-form field strength. Finally, the reason we choose $n^{-b}$ as expansion parameter is because the backreaction corrections of the O6 planes are suppressed precisely with that factor, as motivated in Appendix \ref{App:GreensFunction}. 

We are interested in solutions where the three 2-tori scale differently with $n$. As explained in Section \ref{sec:weak}, we consider $c\ge b>0$. For $c>b$, the second 2-torus is the largest one, followed by the third and the first torus. Note that the opposite regime $c<b$ corresponds to a relabelling of the 2-tori so that we can neglect it. In the special case $b=c$, the second and third tori scale the same and the first torus is parametrically smaller.
This particular scaling is our focus in Section \ref{sec:purespinorloc}, where we discuss an alternative localisation in pure-spinor language, mimicking \cite{Marchesano:2020qvg}. In the present section, we proceed with $c\ge b$.

There is one more crucial assumption we will use in the following. Since our twisted torus is highly anisotropic at large $n$, we expect that the backreaction of the O6 sources effectively only generates field profiles along the second 2-torus (or, in the case $b=c$, the second and third 2-tori). On the other hand, any field profile along the smaller 2-tori is expected to become extremely small at large $n$ and thus be invisible at the first order we consider here. This assumption can be motivated by considering the backreaction in simple systems such as a field satisfying a Poisson equation on a 3D torus or on a very thin cylinder. Studying such toy models, one finds that, at distances from the source that are larger than the sizes of the small cycles, the backreaction becomes effectively 1D, i.e., the field only acquires a profile along the largest cycle, up to non-perturbatively small corrections (see Appendix \ref{App:GreensFunction} for a detailed discussion). We expect the same behaviour to arise in the large-$n$ limit of our more complicated PDE system. We will therefore assume that the first-order coefficients are functions of $y^2$, $y^4$ (or, for $b=c$, functions of $y^2$, $y^3$, $y^4$, $y^5$) and set all other derivatives to zero in the relevant equations. Likewise, all 3D delta sources $\delta(\pi_i)$ appearing in the equations effectively become 1D (or 2D) delta functions in this regime.

\subsection{Solution}

We are now ready to substitute the above Ansatz into the equations of motion (see Appendix \ref{appEOM}) and expand in $1/n$. The $F_2$ Bianchi identity is of particular importance and yields
\begin{align}
\d \left( F_{2,1}^{(0)} + F_{2,2}^{(1)}n^{c-b} + F_{2,3}^{(1)} + F_{2,12}^{(1)}n^{(c-b)/2} + F_{2,23}^{(1)}n^{(c-b)/2} +
F_{2,31}^{(1)} \right) &= - 2\sum_i \delta_{i3}, \label{loc1}
\end{align}
where we used that  $\d F_{2,2}^{(0)}=\d F_{2,3}^{(0)}=0$.\footnote{Recall from Section \ref{sec:weak} that $F_{2,2}^{(0)} \sim e^{24}$, $F_{2,3}^{(0)} \sim e^{35}$ with $\d e^2=\d e^3=\d e^4=\d e^5=0$.} 
We will see below that $\d F_{2,2}^{(1)} = 0$, $F_{2,12}^{(1)} = F_{2,23}^{(1)}=0$ unless $b=c$.
The smeared Bianchi identity \eqref{eq:smear0} furthermore implies $\d F_{2,1}^{(0)}=-2 \sum_i j_{i3}$. Hence, \eqref{loc1} simplifies to
\be \label{loc2}
\d \left(F_{2,2}^{(1)}+ F_{2,3}^{(1)} + F_{2,12}^{(1)} + F_{2,23}^{(1)}+ F_{2,31}^{(1)} \right) = -2\sum_i( \delta_{i3}- j_{i3}).
\ee

The $F_2$ equation of motion only restricts $F_2$ to be co-closed, which will be satisfied in our solution.
Furthermore, the $F_6$ field equations imply that the leading corrections to $F_6$ are a harmonic term plus a term depending on $w^{(1)}$ and $g_{mn}^{(1)}$. We do not spell these corrections out here as they will not play a role in the remainder of this paper.

The Einstein and dilaton equations yield at leading order:
\begin{align}
0 &= 12\frac{\mathcal{T}^{(0)2}}{w^{(0)2}} + 4 \frac{\mathcal{T}^{(0)2}}{w^{(0)}} \nabla^2 w^{(1)} + \mathcal{T}^{(0)} \nabla^2\mathcal{T}^{(1)}
-\sum_{q=2,6} \frac{q-1}{4}  |F_q^{(0)}|^2 + \frac{1}{2} \mathcal{T}^{(0)} \sum_i\delta(\pi_i), \label{loc4} \\
0 &= -\mathcal{T}^{(0)2} R^{(0)}_{mn} -\mathcal{T}^{(0)2} R^{(1)}_{mn} + 4\frac{\mathcal{T}^{(0)2}}{w^{(0)}} \nabla_m \partial_n w^{(1)} + \frac{1}{4} g_{mn}^{(0)} \mathcal{T}^{(0)} \nabla^2\mathcal{T}^{(1)} + 2 \mathcal{T}^{(0)} \nabla_m \partial_n \mathcal{T}^{(1)} \nl
+\frac{1}{2} \sum_{q=2,6} \left( |F^{(0)}_q|_{mn}^2-\frac{q-1}{8}g_{mn}^{(0)} |F^{(0)}_q|^2 \right) + \sum_i \left( \Pi^{(0)}_{i,mn}- \frac{7}{8}g^{(0)}_{mn}\right) \mathcal{T}^{(0)} \delta(\pi_i), \label{loc5} \\
0 &= - 8 \nabla^2 \mathcal{T}^{(1)}- 24\frac{\mathcal{T}^{(0)}}{w^{(0)2}} - 16 \frac{\mathcal{T}^{(0)}}{w^{(0)}} \nabla^2 w^{(1)} + 2\mathcal{T}^{(0)} R_{mn}^{(0)}g^{(0)mn}
\nl + 2\mathcal{T}^{(0)} R_{mn}^{(1)}g^{(0)mn} + 2\sum_i \delta(\pi_i)\,, \label{loc3}
\end{align}
where $R^{(0)}_{mn}$ denotes the Ricci curvature of the smeared metric (with the $n$ scaling taken out) and
\begin{align}
R^{(1)}_{mn} &= - \frac{1}{2} g^{(0)rs} \nabla_m \nabla_n g^{(1)}_{rs}+\frac{1}{2} g^{(0)rs} \left( \nabla_s \nabla_m g^{(1)}_{rn}+\nabla_s \nabla_n g^{(1)}_{rm}\right) -\frac{1}{2} \nabla^2 g_{mn}^{(1)}\,. \label{r}
\end{align}
As in \cite{Junghans:2020acz}, we do not display a superscript ``(0)'' on covariant derivatives and source terms to avoid cluttering the equations with too many indices. In particular, $\nabla_m\equiv\nabla_m^{(0)}$ and $\nabla^2\equiv g^{mn(0)}\nabla_m^{(0)}\nabla_n^{(0)}$.
In deriving the above equations, we used our assumption that the first-order coefficients $w^{(1)}$, $\mathcal{T}^{(1)}$ and $g_{mn}^{(1)}$ only depend on the coordinates $y^2$, $y^4$ of the largest 2-torus (or, in the special case $b=c$, on the coordinates of the two large 2-tori). Furthermore, we used the fact that, in the smeared solution, the warp factor, the dilaton and the internal metric are (covariantly) constant so that $w^{(0)}$, $\mathcal{T}^{(0)}$ and $g_{mn}^{(0)}$ do not appear under derivatives.

In order to simplify the above equations, we now use the smeared equations \eqref{eq:smear1}-\eqref{eq:smear3} to substitute the terms labelled by ``(0)'' by the smeared sources $j_{\pi_i}$. Equations~\eqref{loc2} and \eqref{loc4}-\eqref{loc3} then simplify to the following Bianchi identity and Poisson equations
\begin{align}
& \d \left( F_{2,2}^{(1)} + F_{2,3}^{(1)} + F_{2,12}^{(1)} + F_{2,23}^{(1)} + F_{2,31}^{(1)} \right) =  2\sum_i (j_{i3}-\delta_{i3}), \label{final-eom1} \\
& \nabla^2 \mathcal{T}^{(1)} = -\frac{3}{2}\sum_i \left( j_{\pi_i}- \delta(\pi_i) \right), \label{final-eom2} \\
& \nabla^2 w^{(1)} = \frac{1}{2} \frac{w^{(0)}}{\mathcal{T}^{(0)}}  \sum_i \left( j_{\pi_i}- \delta(\pi_i) \right), \label{final-eom3}
\end{align}
and 
\be
\mathcal{T}^{(0)} R^{(1)}_{mn} - 4 \frac{\mathcal{T}^{(0)}}{w^{(0)}}\nabla_m \partial_n w^{(1)} - 2 \nabla_m \partial_n \mathcal{T}^{(1)} = \sum_i \left( \frac{1}{2}g^{(0)}_{mn}-\Pi^{(0)}_{i,mn}\right)  \left( j_{\pi_i}- \delta(\pi_i) \right). \label{final-eom4}
\ee
Since the equations at this order are linear, the solution with several (intersecting) sources is just the sum of several solutions with one source. For such a single-source solution, we can then make the Ansatz
\begin{align}
\label{eq:linear_solutions}
 w^{(1)}&= w^{(0)} \beta_i\,, & \mathcal{T}^{(1)}&= -3\mathcal{T}^{(0)} \beta_i\,,\\  
\label{eq:linear_solutions2}
g_{mn\parallel}^{(1)}&=2g_{mn}^{(0)}\beta_i\,, & g_{mn\perp}^{(1)}&=-2g_{mn}^{(0)}\beta_i\,, \\
F_{2,12}^{(1)}+F_{2,31}^{(1)}&=4\mathcal{T}^{(0)}\star_{3\perp}^{(0)}\d \beta_i\,, &&
\end{align}
where $\beta_i$ is a scalar function of the coordinate(s) orthogonal to the $i$-th O6 plane. By $g_{mn\parallel}$ and $g_{mn\perp}$ we denote the components of the metric parallel and orthogonal to the source, respectively. This Ansatz is inspired by the backreaction of orientifolds in no-scale Minkowski solutions \cite{Blaback:2010sj}. However, this Ansatz is not yet sufficient to solve the $F_2$ Bianchi identity and the $F_2$ equation of motion at linear order.
Assuming that the leading terms in our expansion are given by the supersymmetric solution of Section \ref{sec:weak}, one finds that the following terms have to be added to $F_2$:
\begin{align}
F_{2,2}^{(1)} + F_{2,3}^{(1)} + F_{2,23}^{(1)} &=   -\mathcal{T}^{(0)} \frac{L_T^{(0)}L_2^{(0)}}{L_3^{(0)}} \left( 4\beta_i
e^{24} + \d \d^\dagger (\xi_i e^{24}) \right) + \mathcal{T}^{(0)}
\frac{L_T^{(0)}L_3^{(0)}}{L_2^{(0)}} \left(4\beta_i e^{35} +  \d \d^\dagger (\xi_i e^{35})\right),
\label{eq:linear_solutions3}
\end{align}
where we define $\d^\dagger=\star_6^{(0)} \d \star_6^{(0)}$ and
$\nabla^2 \xi_i = 10 \beta_i$. Analogous expressions can be derived if the leading terms are non-supersymmetric.
Further note that off-diagonal metric components (with each index on a different 2-torus) are sourced at the order $n^{(a-3b-c)/2}$ by the corresponding off-diagonal Einstein equations. We do not spell out these additional metric corrections here, as they are not relevant for our analysis.

Substituting \eqref{eq:linear_solutions}-\eqref{eq:linear_solutions3} in the above equations of motion, we find that all equations reduce to a single Poisson equation
\begin{equation}
\nabla^2 \beta_i = \frac{1}{2\mathcal{T}^{(0)}} \left( j_{\pi_i}- \delta(\pi_i) \right).
\end{equation}
In our setup, we have 32 intersecting O6 planes (with 8 parallel O-planes in each direction, see Appendix \ref{App:GreensFunction}). The result for all intersecting O6 planes is thus given by the sum of 32 Green's functions $\beta_i$, which have support on the spaces orthogonal to the $i$-th source.
For $b<c$, these spaces are effectively 1D.
For example, let us label with $i=1$ one of the O6 planes with volume form $e^{123}$. We then have $\nabla^2= g^{44(0)} (\partial_4)^2$ and $\delta(\pi_1)\sim \delta(y^4)$. The solution in the range $y^4\in[-1,1]$ is, up to an $\mathcal{O}(1)$ prefactor,
\begin{equation}
\beta_1(y^4) \sim \frac{(y^4)^2}{2}-|y^4|. \label{xx}
\end{equation}
In the special case $b=c$, we would instead have $\nabla^2= g^{44(0)} (\partial_4)^2+g^{55(0)} (\partial_5)^2$ and $\delta(\pi_1)\sim \delta(y^4)\delta(y^5)$. The function $\beta_1(y^4,y^5)$ would then be given by the familiar Green's function on the 2D torus, i.e., the log of a Jacobi theta function.

\section{Lift to 11D}
\label{lift}

The purpose of this section is to lift the previously found 10D strongly coupled solution to 11D. We will give evidence for the existence of 11D geometries which allow a separation of scales without sources but are classically singular. Interestingly, taking into account the corrections coming from the first-order localisation in 10D will be crucial to get the correct sign of the internal curvature of the 7D manifold.

\subsection{11D solution}

The 11D bosonic supergravity equations of motion are\footnote{We have not displayed curvature corrections here, which might be relevant for the resolution of the O6 sources in the 11D uplift.}
\begin{align}
& \hat R_{zz} - \frac{1}{2}\hat g_{zz}\hat R = \frac{1}{2}|\hat G_4|^2_{zz}-\frac{1}{4}\hat g_{zz} |\hat G_4|^2\,, \label{11deinstein1} \\
& \hat R_{Mz} = \frac{1}{2}|\hat G_4|^2_{Mz}\,, \\
& \hat R_{MN} - \frac{1}{2}\hat g_{MN}\hat R = \frac{1}{2}|\hat G_4|^2_{MN}-\frac{1}{4}\hat g_{MN} |\hat G_4|^2\,, \label{11deinstein2} \\
& \d \hat \star_{11} G_4 - \frac{1}{2}G_4 \w G_4 = 0\,.
\end{align}
The hat is used to denote 11D quantities. We choose to work in a vielbein basis with circle fiber $e^z=\d z+C_1$.
Dimensionally reducing these equations on the M-theory circle gives rise to the IIA equations. We can establish the correspondence between the 11D and 10D equations explicitly by writing the 11D metric as
\begin{align}
&\d s_{11}^2 =  \mathcal{T}^{2/3} \d s_{10}^2\, + \mathcal{T}^{-4/3} \left( \d z + C_1\right)^2. \label{eq:UpliftMetric}
\end{align}
Here $z$ is the coordinate of the M-theory circle, $\mathcal{T}=\e^{-\phi}$ is the IIA dilaton, $C_1$ is the RR 1-form potential and $\d s_{10}^2$ is the IIA string-frame metric given by \eqref{IIAmetric}. The M-theory 4-form flux is related to the IIA fluxes as $(G_4)_{MNRP}=(F_4)_{MNRP}$ and $(G_4)_{MNRz}=(H_3)_{MNR}$.
In the absence of $H_3$, we thus have
\begin{equation}
G_4 = F_4,
\end{equation}
where $F_4$ can have both spacetime-filling and internal components. As in the previous sections, we assume that the internal components of $F_4$ vanish and choose to express external $F_4$ flux in terms of its dual internal $F_6$ flux.

We stress that the 11D lift is only possible if one takes into account the backreaction corrections computed in Section \ref{sec:locIIA}. Indeed, one cannot uplift the smeared solution since $F_2$ is not closed in that case and cannot be written in terms of a potential $C_1$. However, the O6 backreaction makes $F_2$ closed, at least away from the localised sources (as implied by the Bianchi identity $\d F_2 = -2 \sum_i \delta_{i3}$). A crucial question is of course what happens at the loci of the delta functions. A natural guess is that, zooming into these regions, the solution will resemble an O6 plane in flat space. Unlike the lift of a D6 brane, the lift of an O6 plane in flat space is not regular. Nonetheless, the singularity in M-theory does not require an explicit source from the viewpoint of the equations of motion and is expected to be resolved to a smooth geometry by quantum effects \cite{Seiberg:1996nz}.
Similarly, quantum/curvature corrections may remove the singularities in our case. However, since our linearised solution is only valid sufficiently far away from the O6 planes, we cannot provide a definite answer here and leave a more detailed analysis of this point for future work.

An interesting observation that may support the above interpretation is the fact that our 11D solution is not warped. The 10D geometries with metrics \eqref{IIAmetric} are warped when the O6 planes are localised.  However, according to the uplift formula \eqref{eq:UpliftMetric}, the uplifted geometry is not warped if $w^2 \sim  \mathcal{T}^{-2/3}$. This relation is obeyed for O6/D6 sources in flat space and for no-scale Minkowski solutions supported by fluxes and O6/D6 sources \cite{Grana:2006kf, Blaback:2010sj}.\footnote{This relation for instance implies that the no-scale vacua supported by $F_2$ flux on generalised CY spaces with O6/D6 sources lift to singular G$_2$-holonomy compactifications of 11D supergravity \cite{Andriolo:2018yrz}.}  Interestingly, our results in Section \ref{sec:locIIA} imply that indeed $w^2 \sim \mathcal{T}^{-2/3}$ at the order of perturbation we work in. Our 11D solution is therefore unwarped at the linear order. It is tempting to take this as further evidence that the 11D solution is sourceless, as proposed above. However, since the IIA solution has multiple intersecting sources, it is not obvious whether such a simple relation between the warp factor and the dilaton still holds when backreaction corrections beyond first order are included.

\subsection{Curvature}

In what follows, we discuss the internal curvature of the 7D manifold. In particular, we will show explicitly that the backreaction corrections computed in Section \ref{sec:locIIA} are crucial to make the curvature consistent with the 11D equations of motion.

We denote the 11D metric with a hat, the 10D string-frame metric without a hat and the unwarped metric with a tilde as before:
$\d s_{11}^2= \hat g_{\mu\nu} \d x^\mu \d x^\nu + \hat g_{mn} e^m e^n + \hat g_{zz}e^z e^z$ with $\hat g_{\mu\nu}= \mathcal{T}^{2/3} g_{\mu\nu} = \mathcal{T}^{2/3} w^2 \tilde g_{\mu\nu}$, $\hat g_{mn}=\mathcal{T}^{2/3} g_{mn}$ and $\hat g_{zz}=\mathcal{T}^{-4/3}$.
Let us also express the 11D curvature tensor in terms of the 10D one:
\begin{align}
\hat R_{zz} &= \frac{2}{3\mathcal{T}^3} \nabla^2 \mathcal{T} + \frac{2}{3\mathcal{T}^4} (\partial \mathcal{T})^2
 +\frac{8}{3w\mathcal{T}^3} (\partial w)(\partial \mathcal{T})
+ \frac{1}{2\mathcal{T}^4}|F_2|^2, \\
\hat R_{\mu\nu} &=  -\frac{w^2}{3\mathcal{T}}\tilde g_{\mu\nu} \nabla^2 \mathcal{T}
-\frac{10w}{3\mathcal{T}} \tilde g_{\mu\nu} (\partial w)(\partial \mathcal{T})
- \frac{w^2}{3\mathcal{T}^2} \tilde g_{\mu\nu}(\partial\mathcal{T})^2 
-3\tilde g_{\mu\nu} (\partial w)^2 -w\tilde g_{\mu\nu} \nabla^2 w \nl  -3 \tilde g_{\mu\nu}, \\
\hat R_{mn} &= -\frac{2}{\mathcal{T}} \nabla_m\partial_n \mathcal{T}
 -\frac{4}{w} \nabla_m\partial_n w
+\frac{2}{\mathcal{T}^2} (\partial_m\mathcal{T})( \partial_n \mathcal{T}) - \frac{1}{3\mathcal{T}}g_{mn} \nabla^2 \mathcal{T}
 - \frac{4}{3w\mathcal{T}}g_{mn} (\partial w)(\partial\mathcal{T}) \nl
- \frac{1}{3\mathcal{T}^2}g_{mn} (\partial \mathcal{T})^2  -\frac{1}{2\mathcal{T}^2}|F_2|^2_{mn} + R_{mn}.
\end{align}
The 7D internal Ricci scalar is:
\begin{align}
\hat R_7 &= -\frac{10}{3\mathcal{T}^{5/3}} \nabla^2 \mathcal{T} + \frac{2}{3\mathcal{T}^{8/3}} (\partial \mathcal{T})^2 
 - \frac{16}{3w\mathcal{T}^{5/3}} (\partial w)(\partial \mathcal{T}) - \frac{4}{w\mathcal{T}^{2/3}} \nabla^2 w 
-\frac{1}{2\mathcal{T}^{8/3}}|F_2|^2 \nl + \frac{1}{\mathcal{T}^{2/3}}R_6. \label{r7}
\end{align}
Substituting the 11D quantities into \eqref{11deinstein1} and \eqref{11deinstein2}, we recover the 10D dilaton and Einstein equations as a consistency check.

The Einstein equations \eqref{11deinstein1}, \eqref{11deinstein2} imply that $\hat R_7$ is positive and $\hat R_4$ is negative:
\begin{equation}
\hat R_7 = \frac{7}{6\mathcal{T}^{8/3}} |F_6|^2 >0, \qquad \hat R_4 = -\frac{4}{3\mathcal{T}^{8/3}} |F_6|^2  <0, \label{11deom}
\end{equation}
where we used that the external $\hat G_4$ satisfies $|\hat G_4|^2=-\mathcal{T}^{-8/3} |F_6|^2$. We further observe that the ratio $\hat R_7/\hat R_4$ is order one, consistent with the no-go theorem of \cite{Gautason:2015tig} reviewed in Section \ref{sec:nogo}. However, we also showed in Section \ref{sec:weak} that the 11D solution is scale-separated. Our solution must therefore circumvent the assumptions of the no-go. In particular, this implies that the KK scale is parametrically decoupled from the internal-curvature scale in our case.

We now turn to the importance of the backreaction corrections. It is instructive to compute $\hat R_7$ for a general uplift with constant warp factor and dilaton.
Using \eqref{r7}, we find
\be
\hat{R}_7 = \frac{R_6}{\mathcal{T}^{2/3}} - \frac{|F_2|^2}{2\mathcal{T}^{8/3}}\,,
\ee
where $R_6$ and $|F_2|^2$ are computed in 10D string frame as before. One can verify that the scalar curvature of the Iwasawa manifold of Section \ref{sec:weak} is negative, $R_6< 0$. Therefore, in the smeared approximation, one falsely concludes $\hat R_7<0$. This is inconsistent with the 11D supergravity equations, which, according to \eqref{11deom}, imply that $\hat R_7$ is positive. Therefore, the corrections induced by backreaction should flip the sign:
\be
\hat{R}_7 = \frac{R_6}{\mathcal{T}^{2/3}} - \frac{|F_2|^2}{2\mathcal{T}^{8/3}} + \textit{backreaction corrections} >0\,,
\ee
where by the first two terms we mean the smeared expressions.
We conclude that the backreaction corrections should be of the order of the negative terms and positive in order to overshoot them. Of course, this is guaranteed by the equations of motion, but nevertheless it is instructive to see how it happens precisely.

Substituting the expansion \eqref{ansatzw}--\eqref{ansatzf} into \eqref{r7}, we find at leading order in $1/n$:
\begin{align}
\hat R_7 &= n^{-a/3-b-c} \left[ \frac{1}{\mathcal{T}^{(0)2/3}}R^{(0)}_{mn}g^{(0)mn} -\frac{1}{2\mathcal{T}^{(0)8/3}}|F_2^{(0)}|^2 + \frac{1}{\mathcal{T}^{(0)2/3}}R^{(1)}_{mn}g^{(0)mn} -\frac{10}{3\mathcal{T}^{(0)5/3}} \nabla^2 \mathcal{T}^{(1)} \right. \nl\qquad\qquad\quad \left.
  - \frac{4}{w^{(0)}\mathcal{T}^{(0)2/3}} \nabla^2 w^{(1)} 
\right]. \label{r7lo}
\end{align}
The corrections (i.e., the last three terms) are derivatives of the 10D dilaton, the warp factor and the internal metric. Such terms are of the same size as the terms that are kept in the smeared limit in the most well-understood compactifications. That the corrections change the sign of $\hat R_7$ might seem in contradiction with the backreaction being small, but this is actually not the case, as explained for instance in \cite{Blaback:2010sj, Junghans:2013xza, Junghans:2020acz}. The up-shot is that corrections to the 4D effective action remain volume and coupling suppressed even though corrections to the 10D/11D equations are relevant.

Upon using the equations of motion in the smeared approximation, together with \eqref{final-eom2}-\eqref{final-eom4} away from the sources, \eqref{r7lo} can be simplified to 
\begin{equation}
    \label{r7final}
\hat R_7 = n^{-a/3-b-c} \frac{7}{6\mathcal{T}^{(0)8/3}}|F_6^{(0)}|^2 > 0.
\end{equation}
We thus see that the 7D internal curvature is manifestly positive, in agreement with \eqref{11deom}, which was obtained by direct computation in 11D. As shown before, we furthermore have $\hat{R}_7/\hat{R}_4$ of order one, consistent with the claims in \cite{Gautason:2015tig}, but the 11D solution is scale-separated, despite $\hat{R}_7/\hat{R}_4$ being order one. This explicitly demonstrates how the assumptions in \cite{Gautason:2015tig} are circumvented.

\section{Comments on the pure-spinor approach}
\label{sec:purespinorloc}

In the previous sections, we localised solutions with orientifold sources to first order in a large-$n$ expansion, in a class of backgrounds comprising the double T-dual of DGKT vacua and following the procedure of \cite{Junghans:2020acz}. This method only relies on the equations of motion and thus captures both supersymmetric and non-supersymmetric solutions. Instead, the work \cite{Marchesano:2020qvg} employed pure-spinor equations to compute the backreaction of supersymmetric DGKT vacua and found that the smeared solution with SU(3) structure is deformed to a localised one with SU(3)$\times$SU(3) structure, up to first order in the expansion parameter $g_s$. In particular, in \cite{Marchesano:2020qvg} it was shown that the localised version of DGKT cannot be embedded into an SU(3) structure. Pure-spinor equations are equivalent to Killing-spinor equations, and thus this second approach captures only supersymmetric solutions.

The complete SU(3)$\times$SU(3)-structure solution of the type IIA supergravity pure-spinor equations can be found for example in \cite{Saracco:2012wc}. However, on top of the given expressions for the fluxes one still has to impose the Bianchi identities, which are not encoded in the pure-spinor equations. Therefore, the strategy in \cite{Marchesano:2020qvg} is to solve the $F_2$ Bianchi identity with an educated Ansatz, while showing at the same time that such an Ansatz is compatible with the general SU(3)$\times$SU(3) solution of \cite{Saracco:2012wc}. 

A posteriori, the first-order correction to $F_2$ found in \cite{Marchesano:2020qvg} can be organised with a simple logic: it amounts to dressing the basis one-forms in $J$ and $\Omega$ with the warp factor associated to the given sources and then plugging the rescaled $J$ and $\Omega$ into the general warped SU(3)-structure expression for $F_2$. Such a localisation prescription was previously suggested in \cite{Grana:2006kf}. For example, when considering a single source, the basis one-forms are corrected by
\begin{align}
\label{eq:warpdressing1}
    &e^m \to  \e^{A} e^m &  &\text{if $e^m$ is parallel to the source},\\
    \label{eq:warpdressing2}
    &e^m \to \e^{-A} e^m &  &\text{if $e^m$ is orthogonal to the source}.
 \end{align}
The generalisation with multiple intersecting sources corresponds to
\begin{equation}
\label{eq:warp_corrected_oneform}
    e^m \to \e^{\sum_i \sign_i(m) A_i} e^m,
\end{equation}
where the sum is performed over the different O6 sources, where $A_i$ is the warp factor associated to the $i$-th source and $\sign_i (m)$ is $+1$ $(-1)$ if $e^m$ is parallel (orthogonal) to the $i$-th source. The total warp factor is then  $\e^{A} = \e^{\sum_i A_i}$.
In practice, all of this implies that in our setup the three-form $\Omega$ receives corrections but the two-form $J$ does not, since it has always one leg parallel and one orthogonal to the source.

The first-order localisation of the DGKT solution forced the authors of \cite{Marchesano:2020qvg} to pass from an SU(3) to an SU(3)$\times$SU(3) structure, because the former cannot accommodate non-trivial warping if the Romans mass is turned on. Besides, with constant warping, the Bianchi identity of $F_2$ with localised sources cannot be satisfied. However, our double T-dual solution belongs to massless type IIA supergravity and thus its localisation might in principle be captured by an SU(3) structure. Below, we give a simple argument showing why this does not seem to be the case and thus even a first-order localisation of the double T-dual of the DGKT solution might require the full SU(3)$\times$SU(3)-structure framework.

Massless type IIA supergravity compactified on a 6D manifold with SU(3) structure allows for non-constant warping \cite{Koerber:2010bx}. The general solution is
\begin{align}
    H_3 &= 0,\\
    g_s F_0 &= 0,\\
    g_s F_2 &= -5 \tilde{m} \e^{-4A} J  - J^{-1}\llcorner \dd \left(\e^{-3A} \Im \Omega \right),\\
    g_s F_4 &= 0,\\
    g_s F_6 &= 3\tilde{m}\e^{-4A}\text{dvol}_6,
\end{align}
where now the constant dilaton is $g_s \equiv \e^{\phi-3A}$ since $\d (\phi-3A)=0$. The operator ${J}^{-1}\llcorner$ means that we construct a bivector out of $J$ and contract it with the form it acts on.\footnote{We define the contraction with a $p$-form $\omega_p$ such that $J^{-1} \llcorner \omega_p = \frac{1}{(p-2)!2} (J^{-1})^{n_1n_2} \omega_{n_1n_2\ldots n_p} e^{n_3...n_p}$.} Warping also corrects the torsion classes, which for our Iwasawa manifold are:
\begin{align}
\label{dJ=Omega}
    &\dd J = 2 \tilde{m} \e^{-A} \Re \Omega,\\
    &\dd \Omega = - \rmi \frac{4}{3}  \tilde{m}\e^{-A} J \wedge J + \mathcal{W}_2 \wedge J + \dd A \wedge \Omega.
\end{align}
When trying to apply the prescription \eqref{eq:warpdressing1}-\eqref{eq:warpdressing2} of dressing the vielbeins we soon encounter a problem. Indeed, $J$ is not corrected while $\Omega$ is, but they both have to satisfy \eqref{dJ=Omega}. Taking for concreteness an O6 plane parallel to $e^{123}$, we find that
\be
\dd J = L_T^2( e^{456}+e^{236}- e^{125}-e^{134}).
\ee
On the other hand, this should be equal to 
\be
2 \tilde{m} \e^{-A} \Re \Omega=	\frac{L_T}{L_2 L_3}\e^{-A} \Re \Omega \rightarrow L_T^2\left[ \e^{-4A} e^{456} + e^{236}-e^{125} - e^{134}  \right].
\ee
Therefore, we see that the term $e^{456}$ does not match unless the warping is trivial, $A \equiv 0$. Even ignoring this issue, in our setup the dressing method of the SU(3)-structure solution results in an unsolvable $F_2$ Bianchi identity at leading order. This leads us to conclude that our AdS$_4$ solution cannot be properly localised within the framework of an SU(3) structure. We expect that a full supersymmetric local solution should be described in the context of SU(3)$\times$SU(3) structure.

Having argued for the need to look at the full SU(3)$\times$SU(3)-structure framework, one can try to follow the approach of \cite{Marchesano:2020qvg} and localise the solutions we are looking at by using pure-spinor equations. The logic is to solve the $F_2$ Bianchi identity with an educated Ansatz, which has to be compatible with the SU(3)$\times$SU(3)-structure expressions of \cite{Saracco:2012wc}. In what follows, we employ the Hodge decomposition to give an expression for $F_2$ that solves the Bianchi identity. A complete match to the SU(3)$\times$SU(3)-structure is more intricate than in \cite{Marchesano:2020qvg} and we leave it for future work.

By exploiting Hodge decomposition, the RR two-form can be written as a sum of co-exact, exact and harmonic pieces, respectively,
\be
\label{eq:F2HodgeDecomposition}
	F_2 = \dd^\dagger \mathcal{K} + \dd \mathcal{C} + F_2^\text{harm}\,.
\ee
We notice that only the co-exact piece contributes to the Bianchi identity 
\be
   \dd \dd^\dagger \mathcal{K} =-2 \sum_i \delta_{i3}.
\ee
Then, the goal is to find an expression for $\mathcal{K}$ in terms of the real two-form $J$ and holomorphic three-form $\Omega$ of the smeared solution, given in \eqref{J}-\eqref{OmegaI}. For simplicity, we work here in the case in which $b = c$ and $L_2 = L_3 = L$. 

Note that all expressions below are meant to hold at linear order in the large-$n$ expansion. Unlike in previous sections, we will adopt a compact notation where we do not spell out the $n$ scalings of the various terms explicitly. The reader should therefore keep in mind that objects like $J$, $\Omega$, $L_i$, etc.~scale non-trivially with $n$ (see Section \ref{sec:weak}). Also note that the Laplacian and the Hodge operators in this section (including implicit ones in $\d^\dagger$) are constructed with the smeared metric and contain a non-trivial scaling as well.

Since we assume the O6 planes to preserve supersymmetry, the delta three-form entering the Bianchi identity has to satisfy the calibration conditions $\sum_i\delta_{i3} \wedge J = 0 = \sum_i\delta_{i3} \wedge \Re \Omega$ \cite{Acharya:2006ne}. These in turn give a constraint on $\mathcal{K}$, namely $\dd \dd^\dagger \mathcal{K} \wedge J = 0 = \dd \dd^\dagger \mathcal{K} \wedge \Re \Omega $. Taking this information into account, we found that the following form of $\mathcal{K}$ solves the Bianchi identity:
\begin{align}
\label{eq:curlyK}
& \mathcal{K} = -\frac{5}{8}g_s^{-1} \Re \Omega + K\,,\\
& K = \frac{{\varphi} }{L_T}\Re \Omega + \Re k  + \dd {\chi } \wedge J\,,
\end{align}
where $k$ is a primitive (2,1)-form. It is such that it satisfies
\begin{equation}
 \frac{{\varphi} }{L_T}\Re \Omega + \Re k = \sum_{i=1}^{32}  4\frac{\varphi_i}{L_T} \text{dvol}_{3\perp,i},
\end{equation} 
where $\varphi_i$ equals, up to a constant factor, the Green's function $\beta_i$ of Section \ref{sec:locIIA} and $\text{dvol}_{3\perp,i}$ is the volume form orthogonal to the $i$-th O6 plane. The function ${\varphi}$ represents a superposition of Green's functions corresponding to each of the 32 O6 planes $({\varphi} = \sum_i \varphi_i)$, whereas ${\chi}$ satisfies 
\be
\nabla^2 \chi =  8 \tilde{m} L_T^{-1} \varphi,
\ee
as a consequence of the calibration conditions. Note that, in our conventions, $\varphi$ and $\chi$ do not scale with $n$.

We also found a one-form
\be
\mathcal{C}_1 = \dd^\dagger({\chi} J)
\ee
such that the coexterior derivative on $K$ can be combined with the exterior derivative on $\mathcal{C}$ to result in
\begin{align}
\label{eq:ddaggerKplusdC1}
\dd^\dagger K + \dd \mathcal{C}_1 =& - J^{-1}\llcorner \dd \left[4L_T^{-1}\varphi \Im \Omega - \star_6 K + \star_6 (\dd \chi \wedge J) \right]
+ 4 L_T^{-1}\varphi J^{-1}\llcorner \dd  \Im \Omega \,.
\end{align}
The logic behind the last step is to let the operator ${J}^{-1}\llcorner$ appear explicitly in the final expression for $F_2$, in order to facilitate the comparison with the solution of the pure-spinor equations of \cite{Saracco:2012wc}. For completeness, we also give the harmonic part of $F_2$:
\be
	F_2^\text{harm} = 3 \tilde m  g_s^{-1} (J_2+J_3),
\ee 
where we are splitting
\begin{align}
	& J \equiv J_1 + J_2 + J_3,   \qquad  J_1 = - L_T^2 e^{16}, \quad J_2 =  - L^2 e^{24}, \quad J_3 = L^2 e^{35}\,.
\end{align}
The equation of motion for $F_2$ (see Eq.~\eqref{eq:App_F2EoM}) in absence of $H_3$ and $F_4$ flux is solved by \eqref{eq:F2HodgeDecomposition} with our $\mathcal{K}$ from \eqref{eq:curlyK} if the one-form $\mathcal{C}$ satisfies
\be
    \mathcal{C} = -\frac{3}{2}\mathcal{C}_1.
\ee
Eventually, the full expression for $F_2$ solving the Bianchi identity and equation of motion for calibrated sources is
\begin{equation}
\begin{aligned}
\label{F2final}
    F_2 &= -5 \tilde mg_s^{-1}  J_1 + 3 \tilde m  g_s^{-1} (J_2+J_3)\\
    & - J^{-1}\llcorner \dd \left[4L_T^{-1}\varphi \Im \Omega - \star_6 K + \star_6 (\dd \chi \wedge J) \right]+ 4 L_T^{-1}\varphi J^{-1}\llcorner \dd  \Im \Omega - \frac{5}{2}\dd \dd^\dagger (\chi J).
    \end{aligned}
\end{equation}
The formulation for $K$ and $\mathcal{C}$ presented here together with Eq.~\eqref{eq:F2HodgeDecomposition} could be useful to match \eqref{F2final} with the SU(3)$\times$SU(3) expression \cite{unpublished:2021}. This is more involved than in \cite{Marchesano:2020qvg} because there the requirement that $F_6$ is vanishing simplifies the problem. We leave this for future work.

\section{Conclusions}
\label{concl}

In this paper, we studied AdS vacua of type IIA string theory and M-theory dual to DGKT \cite{DeWolfe:2005uu}.
An interesting property of such vacua is a parametric scale separation between the AdS curvature scale and the KK scale. These vacua are thus potential counter-examples to swampland conjectures forbidding such a behaviour, in particular to the strong form of the AdS distance conjecture (ADC) \cite{Lust:2019zwm}.
In addition, a common criticism in the literature is that the DGKT solution requires smeared sources and non-zero Romans mass.
Our main result is the construction of solutions without these two requirements.

Our starting point was a family of smeared AdS solutions of (massless) type IIA \cite{Caviezel:2008ik} which are related to DGKT by two (formal) T-dualities.
We then computed the first-order O-plane backreaction in these solutions using the technique of \cite{Junghans:2020acz}.
This approximation turned out to be valid in two interesting regimes with parametrically small backreaction: a weakly coupled regime admitting a family of solutions in perturbative type IIA and a strongly coupled regime with a family of solutions that can be lifted to M-theory. We explicitly constructed the lifted solutions and verified that they indeed exhibit parametric scale separation.

Note that the earlier work \cite{Banks:2006hg} already attempted to construct IIA/M-theory duals of DGKT. However, the authors concluded that neither a weakly coupled IIA nor a strongly coupled M-theory description is uniformly valid in regimes with scale separation.
Let us explain how our analysis differs from the approach of \cite{Banks:2006hg}.
First of all, we considered more general scalings of the fluxes than in \cite{Banks:2006hg}. This allowed us to identify the regimes in which a consistent weakly coupled IIA or a strongly coupled M-theory description is indeed possible.
A second point is that \cite{Banks:2006hg} attempted to lift a smeared IIA solution rather than a properly backreacted one.
The RR field strength $F_2$ is not closed in the smeared solution such that the familiar IIA/M-theory dictionary cannot be used. This problem was circumvented in \cite{Banks:2006hg} by considering a modified expression for $F_2$ which was chosen by hand to satisfy the properly localised Bianchi identity.
However, this ignores that the O6 planes also backreact on all other fields. We have seen that taking into account these backreaction effects is crucial in order to be able to solve the 10D and 11D equations of motion.

Although our results suggest that type IIA and M-theory admit scale-separated AdS vacua, several open questions remain. In particular, our first-order 10D solution is only reliable in regions of the compact manifold where the O-plane backreaction is small. In the large-volume regime we considered, this is true almost everywhere on the manifold except at parametrically small, sub-stringy distances to the O-plane sources where the curvature and the dilaton diverge and string corrections blow up. It would be important to understand better what happens in this near-source region. On physical grounds, it is natural to expect that, zooming into this region, the usual solution for an O6 plane in flat space is recovered. If this is true, the 11D solution should resemble the Atiyah-Hitchin manifold \cite{Seiberg:1996nz} in the regions where our first-order result is not reliable. Our first-order 11D solution would then correspond to the large-volume approximation of a smooth, sourceless M-theory geometry with scale separation.
A possible worry with this interpretation is that problems might arise at the loci where several O6 planes intersect. It would certainly be important to gain more insight into this region. A first step in this direction could be to compute higher-order corrections in the backreaction. We leave a detailed analysis of these issues for future work.

Another point worth mentioning is that our construction also provides the M-theory duals for the non-supersymmetric solutions of \cite{Camara:2005dc, Narayan:2010em, Marchesano:2019hfb}, as they arise in the same class of compactifications that yields the DGKT vacua. Our results thus suggest that M-theory also admits non-supersymmetric AdS vacua with scale separation. However, contrary to the supersymmetric solutions, it is a priori not clear whether the non-supersymmetric ones are stable, and it would be interesting to analyse this further.

Finally, a long-term goal is to determine in general the necessary and sufficient conditions under which string theory admits scale-separated AdS vacua.
We pointed out an apparent obstruction to such vacua in type IIB orientifolds, revising earlier statements in the literature. It would be interesting to see whether our arguments can be generalised to a full no-go theorem in type IIB.
One may also ask how the strong ADC has to be modified if our IIA/M-theory solutions are indeed counterexamples. As discussed earlier, our solutions do satisfy the refined version of the ADC \cite{ Buratti:2020kda} but other solutions seem to violate it \cite{Farakos:2020phe}.
The correct formulation of the ADC is therefore an open question.
It would certainly be important to analyse this further, e.g., using asymptotic scaling symmetries \cite{Junghans:2020acz}
or from a holographic perspective \cite{Alday:2019qrf, Polchinski:2009ch}.
We hope to come back to some of these issues in the future.

\section*{Acknowledgements}
We thank C.~Roupec for initial collaboration and D.~Andriot, A.~Tomasiello and G.~Zoccarato for useful discussions. 
The work of NC is supported by an FWF grant with the number P 30265. VVH is supported by grant nr. 1185120N of the Research Foundation - Flanders (FWO) and thanks Uppsala University for its hospitality during the final stages of this work.
The work of TVR is supported by the KU Leuven C1 grant ZKD1118C16/16/005 and the FWO fellowship for sabbatical research.
The work of TW is supported in part by the NSF grant PHY-2013988.

\appendix
\section{Volume versus curvature scale: examples}\label{examples}

Let us study 5 examples of compact curved spaces with both signs of the curvature. However, we have to keep in mind that for our purposes we are eventually interested in compact spaces which have an everywhere positive Ricci scalar. Since the volume is not always a good measure of the KK scale, we also introduce $L_V$ defined as $\vol_d =  L_V^d$ for a $d$-dimensional space. 
\begin{enumerate}
\item  Consider a product of two 2-spheres with different radii $L_1$ and $L_2$. Then we have
\be
\frac{L^2_{V}}{L_R^2} \sim \frac{L_1}{L_2} + \frac{L_2}{L_1} >1\,.
\ee
So, decoupling is possible but not in the direction we are interested in. We cannot shrink $L_V$ at fixed $L_R$.

\item Consider the orbifold $S^n/\mathbb{Z}_k$. The curvature is $k$-independent, but the volume scales as $1/k$. So, at fixed curvature, we can obtain small volume by increasing $k$. Unfortunately, here the volume is a bad measure of the KK scale since the orbifolding only affects the degeneracies of eigenmodes of the Laplacian (see, e.g., \cite{Polchinski:2009ch}). The KK scale does therefore not decouple from the curvature scale.

\item  Consider Riemann surfaces with genus $g>1$. For the fixed curvature $R=-1$, we have that the volume $L_{V}^2$ is given by
\be
L^2_{V} \sim 2(g-1)\,.
\ee
Again, the decoupling is in the wrong direction and we cannot shrink the volume at fixed curvature radius.

\item   Consider 3D nilmanifolds: The Laplacian and KK spectrum for them was discussed in \cite{Andriot:2016rdd, Andriot:2018tmb}.  As a group manifold, they have one non-zero structure constant, which we take to be $f^{1}{}_{23}$. The structure constants are discretely quantised in terms of an integer $N$. The curvature of such a manifold is given by
\be
R = - \frac{1}{4}L_1^2L_2^{-2}L_3^{-2}(f^{1}{}_{23})^2\,,
\ee
where $L_1,L_2,L_3$ are the sizes of the 3 radii such that $L_{V} = (L_1L_2L_3)^{1/3}$. We then find
\be
\frac{L_{V}}{L_R}  \sim \frac{L_1^2}{L_V^2}|N|\,.
\ee
Increasing $|N|$ does not improve scale separation, so we keep $N$ fixed.  By taking $L_1^2$ much smaller than $L^2_V$, we can however obtain a separation. In this case, the manifold scales non-isotropically but it can be done in such a way that the KK modes associated to the separate circles remain heavy at fixed curvature.\footnote{As an example, take $N=1$, $L_R=1$ and $L_2=L_3=L$. Then  $L_V^3\sim L_1^2\sim L^4$. So, as $L_V$ shrinks, so do $L_1$, $L_2$ and $L_3$.  } 
\item Consider a $d$-dimensional torus with a metric that is conformal to the flat torus:
\be
\d s^2 =  \e^{2B(x)}\left[\sum_{i=1}^d\d x_i^2\right],
\ee
where $x_i\sim x_i+1$. If we take a conformal factor of the form
\be
\e^{-B} =  L^{-1} + \epsilon f(x)\,, 
\ee
with $f$ a periodic function, then as long as $|\epsilon f| < L^{-1}$ this metric is positive everywhere. The Ricci tensor and its integral equal
\begin{align} 
R &=  (2d-2)\e^{-B}\partial^2 \e^{-B}  - d (d-1)(\partial \e^{-B})^2\,, \label{Ricci}\\
\int \d^dx \sqrt{g}R &= \int \d^dx (d-1)(d-2)\e^{dB}(\partial \e^{-B})^2\,. \label{Ricci2}
\end{align}
We see that, for $\epsilon \ll L^{-1}$, $L_R^{-2}$ is proportional to $\epsilon^2$ and therefore $L_V^2/L_R^2\sim \epsilon^2L^2 \ll 1$.
Clearly, we can satisfy $\epsilon \ll L^{-1}$ while keeping $\epsilon$ fixed and making $L$ small. This means that we can shrink the volume at fixed curvature scale. However, while this manifold might have positive integrated Ricci scalar, it is not positive everywhere. To see this, it is enough to realise that $\e^{-B}$ will have local maxima, and one can see from equation (\ref{Ricci}) that the Ricci scalar is negative around those points.
\end{enumerate}

\section{Equations of motion and Bianchi identities of IIA supergravity}
\label{appEOM}

We follow the conventions of \cite{Junghans:2020acz}. The RR and NSNS field equations in string frame are
\begin{align}
    0&=\d (\star_{10} F_2) + H_3 \wedge \star_{10} F_4, \label{eq:App_F2EoM}\\
     0&=\d (\star_{10} F_4) + H_3 \wedge \star_{10} F_6,\\
      0&=\d (\star_{10} F_6),\\
       0&=\d (\mathcal{T}^2\star_{10} H_3) +\star_{10} F_2 \wedge F_0 + \star_{10}F_4 \wedge F_2 + \star_{10}F_6 \wedge F_4,
\end{align}
where $\mathcal{T} \equiv \e^{-\phi}$ and the Hodge star is defined with respect to the full 10D metric, including the warp factor. The Bianchi identities are
\begin{align}
    \d F_0&=0,\\
    \d F_2&=H_3 \wedge F_0 - 2 \sum_i \delta_{i3},\\
    \d F_4&=H_3 \wedge F_2,\\
    \d F_6&=0,\\
    \d H_3&=0.
\end{align}
The Einstein and dilaton equations are 
\begin{align}
    0&=12 \frac{\mathcal{T}^2}{w^2} (1+ (\partial w)^2) + 4 \frac{\mathcal{T}^2}{w} \nabla^2 w + 12 \frac{\mathcal{T}}{w}(\partial w) (\partial \mathcal{T}) + \mathcal{T}\nabla^2 \mathcal{T} + (\partial \mathcal{T})^2 - \frac12 \mathcal{T}^2 |H_3|^2\nl
     -\sum_{q=0}^6\frac{q-1}{4}|F_q|^2 + \frac12 \mathcal{T} \sum_i \delta(\pi_i),\\
    0&=-\mathcal{T}^2 R_{mn} + 4 \frac{\mathcal{T}^2}{w}\nabla_m\partial_n w + \frac{\mathcal{T}}{w}g_{mn}(\partial w)(\partial\mathcal{T}) + \frac14 g_{mn } \mathcal{T}\nabla^2 \mathcal{T} + \frac14 g_{mn}(\partial\mathcal{T})^2 + 2 \mathcal{T}\nabla_m\partial_n \mathcal{T}\nl
    -2 (\partial_m \mathcal{T})(\partial_n \mathcal{T}) + \frac12 \mathcal{T}^2\left(|H_3|^2_{mn}-\frac14 g_{mn}|H_3|^2\right)+\frac12 \sum_{q=0}^6 \left(|F_q|^2_{mn} - \frac{q-1}{8}g_{mn}|F_q|^2\right)\nl
    +\sum_i\left(\Pi_{i,mn}-\frac78 g_{mn}\right)\mathcal{T}\delta(\pi_i),\\
    0&=-8 \nabla^2 \mathcal{T}-24 \frac{\mathcal{T}}{w^2}-\frac{32}{w}(\partial w)(\partial \mathcal{T}) - 24 \frac{\mathcal{T}}{w^2}(\partial w)^2 - 16 \frac{\mathcal{T}}{w}\nabla^2 w + 2 \mathcal{T} R_{mn}g^{mn}-\mathcal{T}|H_3|^2 \nl
    +2 \sum_i \delta(\pi_i),
\end{align}
where $w \equiv L_H \e^A$ and the stress-energy of the $i$th O-plane is proportional to the projector
\begin{equation}
    \Pi_{i,mn} = -\frac{2}{\sqrt{g_{\pi_i}}}\frac{\delta \sqrt{g_{\pi_i}}}{\delta g^{mn}} = (g_{\pi_i})^{\alpha \beta}\frac{\partial y^l}{\partial \xi^\alpha_i}\frac{\partial y^p}{\partial \xi_i^\beta}g_{ml}g_{np}.
\end{equation}
We denote by $\delta(\pi_i)$ the delta distribution with support on the (torsional) 3-cycle $\pi_i$ wrapped by the $i$-th O6 plane and by $\delta_{i3}$ the corresponding 3-form that integrates to one over the dual chain $\tilde \pi_i$. We define $\delta(\pi_i) \equiv \frac{\sqrt{g_{\pi_i}}}{\sqrt{g_6}}\delta^{(3)}(y)$ and $\int_{\tilde \pi_i}\delta_{i3}\equiv \int_{\tilde \pi_i} \d^3 y \delta^{(3)}(y)=1$ in local coordinates such that
\begin{equation}
    \int_{\pi_i} {\rm dvol}_{\pi_i} = \int {\rm dvol}_{\pi_i }\wedge \delta_{i3} = \int \d^6 y \sqrt{g_6} \delta(\pi_i).
\end{equation}
We also set $g_6 \equiv {\rm det}(g_{mn})$ and $g_{\pi_i} \equiv {\rm det}((g_{\pi_i})_{\alpha\beta})$ with world-volume metric $(g_{\pi_i})_{\alpha \beta} = g_{mn} \frac{\partial y^m}{\partial \xi^\alpha_i}\frac{\partial y^n}{\partial \xi^\beta_i}$ and world-volume coordinates $\xi^\alpha_i$, $\alpha=1,2,3$.

The smeared approximation amounts to replacing the source terms by
\begin{equation}
    \delta (\pi_i) \to j_{\pi_i} = \frac{\text{vol}_{\pi_i}}{\text{vol}_6},\qquad \delta_{i3} \to j_{i3} = \frac{{\rm dvol}_{\tilde \pi_i}}{\text{vol}_{\tilde \pi_i}},
\end{equation}
where 
\begin{equation}
    \text{vol}_6 = \int \d^6 y \sqrt{g_6}, \quad \text{vol}_{\pi_i} = \int_{\pi_i} \d^3 y \sqrt{g_{\pi_i}}, \quad \text{vol}_{\tilde \pi_i} = \int_{\tilde \pi_i} \d^3 y \sqrt{g_{\tilde \pi_i}},
\end{equation}
such that $\int_{\tilde \pi_i} \delta_{i3} = \int_{\tilde \pi_i}j_{i3}=1$.

\section{Kaluza-Klein spectrum on a 6D/7D twisted torus}
\label{appKK}

In this appendix, we would like to investigate the effects of twisting on the Kaluza-Klein spectrum. We will see that it is reasonable to assume that twisting does not lower the KK scale with respect to the flat-torus case.

We would like to generalise the formulae of \cite{Andriot:2016rdd} valid for a 3D nilmanifold to our case of a 6D twisted torus with four different geometric fluxes.\footnote{Note that the 
formulae of \cite{Andriot:2016rdd} do not immediately apply to our case, as we do not have a product of two 3D nilmanifolds.}
To be general, let us consider the metric
\begin{equation}
\d s_6^2 = \sum_a r_a^2 e^a e^a \label{m1}
\end{equation}
with radii $r_a$. The vielbeins satisfy $\d e^1 = -f e^{23}- g e^{45}$, $\d e^6 = -h e^{25}- j e^{34}$ with $f,g,h,j\in \mathbb{Z}$ and $\d e^a =0$ for $a=2,3,4,5$. This reduces to our Ansatz in Section \ref{sec:weak} for $r_1=r_6=L_T$, $r_2=r_4=L_2$, $r_3=r_5=L_3$, $f=g=h=j=1$.

We choose the parametrisation
\begin{equation}
e^1 = \d y^1 - f y^2 \d y^3 - g y^4 \d y^5, \qquad e^6 = \d y^6 + h y^5 \d y^2 - j y^3 \d y^4. \label{m2}
\end{equation}
The Laplacian in the coordinate basis is thus
\begin{equation}
\Delta = \Delta_\text{torus} + \Delta_f + \Delta_g+ \Delta_h+ \Delta_j \label{lap}
\end{equation}
with $\Delta_\text{torus}= \sum_m r_m^{-2}\partial_m^2$ and
\begin{align}
\Delta_f &= r_3^{-2}f^2(y^2)^2\partial_1^2 + 2r_3^{-2}fy^2\partial_1\partial_3, \\ \Delta_g &= r_5^{-2}g^2(y^4)^2 \partial_1^2 + 2r_5^{-2}gy^4\partial_1\partial_5, \\ \Delta_h &= r_2^{-2}h^2(y^5)^2\partial_6^2 - 2r_2^{-2}hy^5\partial_2\partial_6, \\ \Delta_j &= r_4^{-2}j^2(y^3)^2 \partial_6^2 + 2r_4^{-2}jy^3\partial_4\partial_6.
\end{align}

Assuming for the moment that $g=h=j=0$, our manifold simplifies to a product of a 3D torus and a 3D nilmanifold, with Laplacian $\Delta_\text{torus} + \Delta_f$. The eigenfunctions are now simply given by the eigenfunctions found in \cite{Andriot:2016rdd} (times the usual exponential factors accounting for the extra 3D torus).
Here, we focus on those modes whose flat-torus analogues would be excitations along the $y^1$ circle, as the masses of these modes receive corrections due to the twisting \cite{Andriot:2016rdd}. The modes excited along the remaining circles are given by the usual exponentials, and their masses do not receive any corrections compared to the flat-torus case \cite{Andriot:2016rdd}. More generally, one could also consider modes which are excited along several circles at once. However, we will not do so, as we expect that such mixed modes are heavier than those we consider.

We denote the eigenfunctions of $\Delta_\text{torus} + \Delta_f$ by $\e^{2\pi i k_1y^1}u_f(y^2,y^3)$, with eigenvalues $-m_f^2$. To avoid cluttering, we suppressed indices $n\in \mathbb{N}$ and $\ell=0,1,\ldots, |k_1|-1$ labeling a degeneracy in the eigenfunctions $u_f$ and a corresponding dependence of $m_f^2$ on $n$ \cite{Andriot:2016rdd}.

We now want to generalise the above to the 6D case with all four geometric fluxes turned on. We denote by $\e^{2\pi i k_1y^1}u_g(y^4,y^5)$, $\e^{2\pi i k_6 y^6}u_h(y^5,y^2)$ and $\e^{2\pi i k_6y^6}u_j(y^3,y^4)$ the corresponding eigenfunctions of $\Delta_\text{torus} + \Delta_{g,h,j}$ with eigenvalues $-m^2_{g,h,j}$. A natural ansatz is then to take the eigenfunctions for the full Laplacian to be products of the individual eigenfunctions on the nilmanifolds without ``double-counting'' the exponential factors $\e^{2\pi i k_1y^1}$ and $\e^{2\pi i k_6y^6}$. Concretely, we consider the following Ansatz for two independent sets of eigenfunctions:
\begin{equation}
U_1=\e^{2\pi i k_1 y^1}u_f(y^2,y^3) u_g(y^4,y^5), \qquad U_6=\e^{2\pi i k_6y^6}u_h(y^5,y^2) u_j(y^3,y^4).
\end{equation}

Acting on $U_1$, $U_6$ with the Laplacian \eqref{lap} and using the properties of the 3D eigenfunctions of \cite{Andriot:2016rdd}, we find that $U_1$, $U_6$ are indeed eigenfunctions:
\begin{align}
\Delta U_1 &= -\left(m_f^2+m_g^2- \frac{(2\pi k_1)^2}{r_1^2}\right)U_1 \nll = -\left( \frac{(2\pi k_1)^2}{r_1^2} + 2\pi|k_1| \left[\frac{(2n_1+1)|f|}{r_2r_3} + \frac{(2m_1+1)|g|}{r_4r_5}\right] \right)U_1, \\
\Delta U_6 &= -\left(m_h^2+m_j^2- \frac{(2\pi k_6)^2}{r_6^2}\right)U_6 \nll = -\left( \frac{(2\pi k_6)^2}{r_6^2} + 2\pi|k_6| \left[\frac{(2n_6+1)|h|}{r_2r_5} + \frac{(2m_6+1)|j|}{r_3r_4}\right] \right)U_6,
\end{align}
where $k_a \in \mathbb{Z}$ and the integers $n_a,m_a\in \mathbb{N}$ label the degeneracy mentioned above.

In the notation of Section \ref{sec:weak}, we thus find the following KK masses:
\begin{align}
M_{\text{KK},k_1,n_1,m_1}^2 &= \frac{(2\pi k_1)^2}{L_T^2} + (n_1+m_1+1) \frac{4\pi|k_1|}{L_2L_3}, \\
M_{\text{KK},k_6,n_6,m_6}^2 &= \frac{(2\pi k_6)^2}{L_T^2} + (n_6+m_6+1) \frac{4\pi|k_6|}{L_2L_3}.
\end{align}
We thus see that the second terms in the above formulae, which come from twisting, increase the KK masses with respect to the flat-torus case.
As stated before, there are also modes on the torus base spanned by $e^2$, $e^3$, $e^4$, $e^5$ with squared masses $M_{\text{KK},k_a}^2=\frac{(2\pi)^2 (k_2^2+k_4^2)}{L_2^2}+\frac{(2\pi)^2 (k_3^2+k_5^2)}{L_3^2}$. Furthermore, modes for which several $k_a$ are non-zero on the fibers and the base are expected to be heavier than those we computed.

Let us finally also consider the twisting of the circle fiber in the 7D case:
\begin{equation}
\d s_7^2 = \sum_a g_s^{-2/3} r_a^2 e^a e^a + g_s^{4/3} (e^z)^2
\end{equation}
with $\d e^z = F_2 = 5\frac{\tilde m}{g_s} L_T^2 e^{16} - 3 \frac{\tilde m}{g_s} L_2^2 e^{24}+3\frac{\tilde m}{g_s} L_3^2 e^{35}+\ldots$ and $\d e^a =0$ otherwise, i.e., we set all 6D metric fluxes to zero, $f=g=h=j=0$. Applying the above logic to the present case then yields
\begin{align}
M_{\text{KK},k_7,n_7,m_7,p_7}^2 &= \frac{(2\pi k_7)^2}{g_s^{4/3}} + \left[5(2n_7+1)+ 3(2m_7+1)+ 3(2p_7+1)\right] \frac{2\pi|k_7| \tilde m}{g_s^{1/3}},
\end{align}
where $k_7\in\mathbb{Z}$ and $n_7,m_7,p_7\in \mathbb{N}$ denote a degeneracy as before.
The correction to the ``naive" result that neglects the non-trivial fibration thus scales like $\frac{\tilde m}{g_s^{1/3}} \sim n^{-1/3}$ in the strongly coupled example. On the other hand, $g_s^{-4/3}\sim n^{-1/12}\gg n^{-1/3}$. The leading contribution to the squared masses of the lowest KK modes on the M-theory circle is therefore $g_s^{-4/3}$ as in the flat case, whereas the corrections due to $F_2$ are parametrically suppressed.

\section{Field profiles on the twisted torus at large $n$}
\label{App:GreensFunction}

In this appendix, we argue that only derivatives with respect to the coordinates of the largest 2-tori (i.e., $y^2$, $y^4$ for $c>b$ and $y^2$, $y^3$, $y^4$, $y^5$ for $c=b$) are relevant in the equations of motion at the order of the localisation procedure we are interested in. We motivate this assumption using two toy examples of backreaction. In particular, we consider point sources on a thin 2D cylinder and on an anisotropic 3-torus. The latter example also motivates our Ansatz with $n^{-b}$ as the expansion parameter.

The first example we consider is a 2D cylinder with metric $\dd s^2 = L_1^2(\dd y^1)^2 + L_2^2(\dd y^2)^2$, where $y^1$ is the circle coordinate and we are interested in the limit $L_1 / L_2 \to 0$. 
The Poisson equation
\be
    \Delta G = \frac{1}{L_1L_2}\delta(y^1)\delta(y^2)
\ee
is solved by (see, e.g., \cite{ooguri})
\begin{align}
    G(y^1,y^2) &= \frac{1}{4\pi} \ln \left|\sin 2\pi\left(y^1 + i \frac{L_2}{L_1}y^2 \right) \right|^2 \notag \\
    &= \frac{1}{4\pi}\ln\left( \sin^2 (2\pi y^1) \cosh^2\left( 2\pi \frac{L_2}{L_1} y^2\right) +\cos^2( 2\pi y^1) \sinh^2 \left(2\pi \frac{L_2}{L_1} y^2\right) \right) .
\end{align}
At distances from the source larger than the circle length, i.e., for $y^2 \gg L_1/L_2$, we have that $\cosh^2\left( 2\pi \frac{L_2}{L_1} y^2\right) \sim \sinh^2\left( 2\pi \frac{L_2}{L_1} y^2\right) \sim \frac{1}{{{4}}}\exp\left( {4} \pi \frac{L_2}{L_1} y^2\right)$, and thus the Green's function approximates to
\be
    G(y^1,y^2) = \frac{L_2}{L_1} y^2 + \text{ const. } + \frac{1-2 \cos^2(2 \pi y^1)}{2\pi}\,\e^{-4\pi L_2 y^2 /L_1} + \mathcal{O}\left(\e^{-8\pi L_2 y^2 /L_1}\right).
\ee
The dependence on $y^1$ is thus exponentially suppressed, and the Green's function is effectively the Green's function of the 1D Laplacian. In the limit $L_1 / L_2 \to 0$, this approximation is valid for all $y^2 >0$.

Analogously, taking the large-$n$ limit in our twisted-torus setup implies $L_T/L_2  \to 0$ and, for the case $c>b$, $L_3/L_2\to 0$. We therefore expect that the O-plane backreaction does not generate any field profiles along the smaller 2-tori at the linear order in the large-$n$ expansion we consider.

As a second example, we consider a source on an anisotropic 3-torus.
We parametrise the torus by $y^4$, $y^5$ and $y^6$ and denote the circle lengths by $L_2$, $L_3$ and $L_T$, which we assume to scale with $n$ as in \eqref{eq:general_scalings}. The scalings are thus analogous to those of the transverse space seen by the O6 plane with volume form $\sim e^{123}$ in our twisted-torus setup.

We want to study the Poisson equation on the 3-torus.
This equation indeed describes the backreaction of an O6 plane in simple setups such as the Minkowski solutions of \cite{Blaback:2010sj}.
The Green's function of the Laplacian on a flat isotropic $d$-dimensional torus was derived in \cite{Shandera:2003gx, Andriot:2019hay} (see also \cite{CourantHilbert} for the case $d=3$). We now briefly sketch the derivation, adapting it to the case of a non-isotropic 3-torus \cite{Junghans:2020acz}.

A naive guess for the Green's function $G(\vec y)$ is
\begin{equation}
    G(\vec y) =  - \frac{1}{L_T L_2 L_3} \sum_{\vec{n} \in \mathbb{Z}^3\backslash\{\vec{0}\}} \frac{\e^{2\pi i \vec{n} \cdot \vec y}}{4 \pi^2 \sum_i (n_i/L_i)^2},
\end{equation}
where $\vec{y}=(y^4,y^5,y^6)$ and $L_i =\{L_2, L_3, L_T\}$. However, this sum is not absolutely convergent. To amend this problem, one uses
\begin{equation}
    \frac{1}{4\pi^2 \sum_i (n_i/L_i)^2 } = \int_0^\infty \d t \,\e^{-4 \pi^2 \sum_i (n_i/L_i)^2 t}
\end{equation}
and then defines the regularised Green's function as the one in which the sum and the integral are interchanged \cite{Shandera:2003gx}. In our case, this gives
\be
    G(\vec{y}) = \frac{1}{L_T L_2 L_3} \int_0^{+\infty} \dd t \left[1 - \theta_3\left(y^4\bigg| \frac{4 \pi it}{L_2^2}\right)\theta_3\left(y^5\bigg| \frac{4 \pi it}{L_3^2}\right)\theta_3\left(y^6\bigg| \frac{4 \pi it}{L_T^2}\right) \right] ,
\ee
where $\theta_3 (y|\tau) = \sum_{n \in \mathbb{Z}} \e^{2 \pi i (n y + \frac{n^2}{2}\tau)}$ is a Jacobi theta function.

We can now determine the leading-order scaling behaviour of the Green's function at large $n$. Since we consider $L_2$ our fastest growing length scale, we perform a change of integration variables by $ u = t/L_2^2 $. The Green's function then becomes
\be
    G(\vec{y}) = \frac{L_2}{L_T L_3} \int_0^{+\infty} \dd u \left[1 - \theta_3\left(y^4\bigg| 4 \pi i u\right)\theta_3\left(y^5\bigg| \frac{4 \pi i u}{(L_3/L_2)^2}\right)\theta_3\left(y^6\bigg| \frac{4 \pi i u}{(L_T/L_2)^2}\right) \right] .
\ee
A numerical analysis shows that, for a fixed point $\vec{y}$ away from the source, the integral approaches a constant in the large-$n$ limit. The non-trivial $n$ scaling of $G$ is thus exclusively in the prefactor $L_2/L_TL_3$. The backreaction of an O-plane in this setting is given by $g_s G$  \cite{Blaback:2010sj} and thus scales like $g_s L_2/(L_T L_3) \sim n^{-b}$. Hence, $n^{-b}$ is the natural expansion parameter for our Ansatz in Section \ref{sec:ansatz}, which will indeed turn out to be self-consistent.

One furthermore verifies that the derivative of the integral along the large circle(s) (i.e., $y^4$ or, in the case $c=b$, $y^4$, $y^5$) also approaches a constant at large $n$, whereas derivatives along the small circle(s) (i.e., $y^5$, $y^6$ or, in the case $c=b$, $y^6$) vanish exponentially in the large-$n$ limit.
We thus again see that derivatives with respect to the smaller circles become negligible in the large-$n$ expansion.

Let us finally show that the backreaction of an O6 plane in our setup only generates field profiles along the \emph{transverse} (rather than the parallel) directions, as assumed in Section \ref{sec:locIIA}. This assumption would of course be correct on an ordinary flat torus. However, the torus we are considering is twisted and thus our assumption requires justification. For concreteness, consider the backreaction generated by the O6 planes along $e^{123}$. The orientifold involution acts as $\sigma(y^1,y^2,y^3,y^4,y^5,y^6) = (y^1,y^2,y^3,-y^4,-y^5,-y^6)$, so that an O6 plane sits at $y^4=y^5=y^6=0$.
The metric in the coordinate basis is given in \eqref{m1}, \eqref{m2}, where $r_1=r_6=L_T$, $r_2=r_4=L_2$, $r_3=r_5=L_3$ and $f=g=h=j=1$ in our setup. In the following, it will be convenient to redefine $y^6\to y^6-y^2y^5$ such that
\begin{equation}
e^1 = \d y^1 - y^2 \d y^3 - y^4 \d y^5, \qquad e^6 = \d y^6 - y^2 \d y^5 - y^3 \d y^4.
\end{equation}
In these coordinates, the metric is consistent with the identifications
\begin{align}
 y^1&\sim y^1+2, \\
 (y^2,y^1,y^6) &\sim \left(y^2+2, y^1 + 2y^3 , y^6 + 2y^5\right),\\
 (y^3,y^6) &\sim \left( y^3+2, y^6 +2y^4 \right),  \\ 
 (y^4,y^1) &\sim \left( y^4+2, y^1 +2y^5 \right),\\
 y^5 &\sim y^5+2,\\
 y^6 &\sim y^6+2.
\end{align}
Using this, we can infer that there are  O-planes at $y^4,y^5,y^6\in\{0,1\}$ (plus an infinite number of images due to the above identifications). Note that we chose the identifications such that the $H_3$-flux number on the T-dual side is an even integer.\footnote{For odd flux numbers, e.g., replacing every $2$ in the above identifications by a $1$, an interesting subtlety arises. Indeed, one can verify that, in this case, there are no fixed points at $y^4=y^5=\frac{1}{2},y^6=\{0,\frac{1}{2}\}$. This means that two O6 planes seem to be ``missing"  compared to the T-dual flat-torus case. It therefore seems that some energy density is lost upon T-dualising, unless some of the O-planes carry a different energy density than usual. The T-dual version of this phenomenon was understood in \cite{Frey:2002hf}, where it was found that a toroidal orientifold with odd $H_3$-flux number is only consistent if the O-planes carry additional localised flux (see also \cite{Gur-Ari:2013sba} for a discussion of the twisted-torus case). In order to avoid such subtleties, we consider even flux  numbers here.}
Ultimately, we are interested in an orientifold of the orbifold $T^6/\mathbb{Z}_2 \times \mathbb{Z}_2$. In this case, we have further images under the $\mathbb{Z}_2 \times \mathbb{Z}_2$ orbifold, as explained in Section \ref{sec:weak}. However, let us ignore these intersecting image O-planes for the moment.

According to the above discussion, the presence of the O6 planes and their parallel images yields delta-function sources of the form $\sim \delta(y^4+\mathbb{Z})\delta(y^5+\mathbb{Z})\delta(y^6+\mathbb{Z})$ and thus only breaks translation invariance along $y^4$, $y^5$ and $y^6$, as it would be the case without the twisting. Neglecting the dependence on the smaller circles as explained above, we conclude that the O-plane backreaction at linear order generates a field profile along $y^4$ for the case $c>b$ and along $y^4$, $y^5$ for the case $c=b$.

In Section \ref{sec:locIIA}, we consider the equations of motion in a vielbein basis. The derivative operators $\nabla_a$ do therefore not coincide with those in the coordinate basis in general. However, using that derivatives with respect to $y^1$ and $y^6$ are negligible, one verifies that the differential operators in the vielbein basis reduce to the corresponding operators without geometric fluxes. In particular,
one can see from \eqref{lap} that the Laplacian on the twisted torus then reduces to the Laplacian on the flat torus. Furthermore, any derivative in the vielbein basis acting on a scalar reduces to an ordinary derivative, i.e., $\nabla_a = e^m_a\nabla_m=\delta^m_a\frac{\partial}{\partial y^m}$. One can verify that the same is true for the operators $\nabla_a\nabla_b$ that appear in the Einstein equations. An exception are some operators $\nabla_a\nabla_b$ with one index transverse and one parallel to the O-planes (e.g., one finds that $\nabla_3\nabla_6 w = \frac{L_T^2}{2L_2^2} \frac{\partial}{\partial y^4} w$ at leading order). However, such operators only appear in the corresponding off-diagonal Einstein equations, which are not relevant for our analysis.

We can therefore consistently neglect all operators $\nabla_a$ in the equations of motion except for $\nabla_4$ (and, for $c=b$, $\nabla_5$).
Indeed, we show in Section \ref{sec:locIIA} that the equations of motion at linear order admit a solution that is consistent with this assumption.

Analogous conclusions apply for the field profiles generated by the other O-planes. For example, the O-planes with current $ \sim e^{236}$ will generate a non-trivial field dependence on $y^2$ for $c>b$ and on $y^2$, $y^3$ for $c=b$. When computing the backreaction corrections generated by these O-planes and their parallel images, we can therefore neglect all derivatives in the equations of motion except for $\nabla_2$ and, possibly, $\nabla_3$. At linear order in the large-$n$ expansion, the total backreaction correction to a given field is then simply the sum of the corrections generated by each O-plane.

\bibliographystyle{JHEP}
\bibliography{refs}

\end{document}